\documentclass[aps,prd,twocolumn,showpacs,preprintnumbers,nofootinbib,amsmath,amssymb]{revtex4}

\usepackage{amsmath}
\usepackage{color}
\usepackage{bm}
\usepackage{dcolumn}
\usepackage{graphicx}
\usepackage{multirow}

\newcommand{\beq}{\begin{eqnarray}}
\newcommand{\eeq}{\end{eqnarray}}

\newcommand{\tr}{\ensuremath{\mathrm{Tr}}}

\newcommand{\ds}{\ensuremath{\displaystyle}}

\def\spose#1{\hbox to 0pt{#1\hss}}
\def\ltapprox{\mathrel{\spose{\lower 3pt\hbox{$\mathchar"218$}}
 \raise 2.0pt\hbox{$\mathchar"13C$}}}

\begin{document}

\title{Curvature of the {chiral} pseudo-critical line in QCD: continuum 
extrapolated results}

\author{Claudio Bonati}
\email{bonati@df.unipi.it}

\author{Massimo D'Elia}
\email{delia@df.unipi.it}

\author{Marco Mariti}
\email{mariti@df.unipi.it}

\author{Michele Mesiti}
\email{mesiti@pi.infn.it}

\author{Francesco Negro}
\email{fnegro@pi.infn.it}
\affiliation{
Dipartimento di Fisica dell'Universit\`a
di Pisa and INFN - Sezione di Pisa,\\ Largo Pontecorvo 3, I-56127 Pisa, Italy}

\author{Francesco Sanfilippo}
\email{f.sanfilippo@soton.ac.uk}
\affiliation{School of Physics and Astronomy, University of Southampton, SO17 1BJ Southampton, 
United Kindgdom}

\date{\today}

\begin{abstract}
We determine the
curvature of the pseudo-critical line of strong interactions
by means of numerical simulations at imaginary chemical potentials.
We consider
$N_f=2+1$ stout improved staggered fermions with physical quark masses and
the tree level Symanzik gauge action,
and explore 
four different sets of lattice spacings, 
corresponding to $N_t = 6,8,10,12$, in order to extrapolate results
to the continuum limit.
Our final estimate is $\kappa = 0.0135(20)$.
\end{abstract}

\maketitle

\section{Introduction}
\label{intro}

The exploration of the phase diagram of strongly interacting matter in the
temperature - baryon chemical potential ($T - \mu_B$) plane is being pursued
both by experimental and by theoretical investigations.  The comparison between
the chemical freeze-out line~\cite{freezeout1, freezeout2, freezeout3,
freezeout4, freezeout5, freezeout6,freezeout7,freezeout8} and the crossover
line, corresponding to chiral symmetry restoration, is one of the main issues.
In principle these two lines are not expected to coincide, however an exact statement
about their interrelation will provide useful information about the dynamics of
strong interactions. That requires a precise determination of both lines.

From the theoretical point of view, Lattice QCD simulations represent the best
first principle tool to provide information about the chiral
transition\footnote{We speak of chiral transition even if present lattice
studies provide evidence for a crossover~\cite{aefks,afks,betal,tchot,tchot2}.}
temperature $T_c$: present results provide consistent evidence for $T_c \simeq
155$ MeV at $\mu_B = 0$.  Unfortunately, as one moves to finite baryon chemical
potential, direct numerical simulations are presently hindered by the so-called
sign problem, stemming from the complex nature of the fermion determinant when
$\mu_B \neq 0$.  However, various methods have been proposed to circumvent the
problem in the regime of small chemical potentials, where the pseudo-critical
line can be well approximated by a quadratic behavior\footnote{{We note that
a possible ambiguity in the denominator of the quadratic term, i.e.
whether we take $\mu_B/T_c(\mu_B)$ or  
$\mu_B/T_c(0)$ as an expansion variable, is irrelevant as long as 
just the quadratic term is considered, since it only affects
higher order terms.}}  
 in $\mu_B^2$:
\begin{equation}\label{corcur}
\frac{T_c(\mu_B)}{T_c}=1-\kappa \left(\frac{\mu_B}{T_c}\right)^2\, +\, 
O(\mu_B^4)\, ,
\end{equation}
where the coefficient $\kappa$ defines the curvature of the pseudo-critical line
$T_c (\mu_B)$. Information about $\kappa$ can be obtained for instance by 
Taylor expansion techniques~\cite{Allton:2002zi, Kaczmarek2011,
Endrodi2011,Borsanyi2012}), by analytic continuation from imaginary chemical 
potentials~\cite{fp1,dl1,adgl,wlc,ceasu2,ccdmp,ccdp,naganaka,nf2_ccdps,Laermann2013,ccp,crow}, by 
reweighting techniques~\cite{FK1, FK2} or
{by a reconstruction of the canonical partition function}~\cite{kf,afhl}.

Recent numerical investigations~\cite{ccp,crow}, adopting the method of
analytic continuation with improved discretizations at or close to the physical
point of $N_f = 2+1$ QCD, have provided results for $\kappa$ which are
generally larger than previous estimates obtained by the Taylor expansion
technique~\cite{Kaczmarek2011,Endrodi2011,Borsanyi2012}.

In particular, in Ref.~\cite{crow} we performed numerical simulations adopting
an improved stout staggered fermion discretization on lattices with $N_t =
6,8$, leading to a preliminary estimate $\kappa \sim 0.013$, to be compared
with previous determinations obtained by Taylor
expansion~\cite{Kaczmarek2011,Endrodi2011,Borsanyi2012}, reporting $\kappa \sim
0.006$.

In the present study we aim at extending our results in two directions.  First,
we increase the number of imaginary chemical potentials explored on lattices with
$N_t = 8$, in order to obtain a better control over the 
analytic continuation systematics and to perform a deeper comparison between
the cases in which a strange quark chemical potential is included or not.  Then
we extend simulations for $\mu_s = 0$ to two new sets of lattice spacings,
corresponding to $N_t = 10$ and $N_t = 12$, in order to perform a continuum
extrapolation of our determination of $\kappa$.  As a byproduct, we also
discuss the behavior of the continuum extrapolated chiral susceptibilities as a
function of $\mu_B$, in order to assess the possible influence of the baryon
chemical potential on the strength of the transition, which is relevant to the
possible existence of a critical endpoint in the $T-\mu_B$ plane.

The paper is organized as follows.  In Section~\ref{sec2} we provide some
details about the lattice discretization adopted in this study, about the
various explored setups of chemical potentials, about the observables chosen to
locate $T_c$ and their renormalization.  In Section~\ref{sec_results} we
discuss our numerical results and finally, in Section~\ref{sec_conclusions}, we
draw our conclusions.

\section{Numerical setup}
\label{sec2}

As in Ref.~\cite{crow}, we consider a lattice discretization of $N_f=2+1$ QCD
in the presence of purely imaginary quark chemical potentials. 
We consider the following euclidean partition function 
\begin{eqnarray}\label{partfunc}
\mathcal{Z} &=& \int \!\mathcal{D}U \,e^{-\mathcal{S}_{Y\!M}} \!\!\!\!\prod_{f=u,\,d,\,s} \!\!\! 
\det{\left({M^{f}_{\textnormal{st}}[U,\mu_{f,I}]}\right)^{1/4}}
\hspace{-0.1cm}, \\
\label{tlsyact}
\mathcal{S}_{Y\!M}&=& - \frac{\beta}{3}\sum_{i, \mu \neq \nu} \left( \frac{5}{6}
W^{1\!\times \! 1}_{i;\,\mu\nu} -
\frac{1}{12} W^{1\!\times \! 2}_{i;\,\mu\nu} \right), \\
\label{fermmatrix}
(M^f_{\textnormal{st}})_{i,\,j}&=&am_f \delta_{i,\,j}+\!\!\sum_{\nu=1}^{4}\frac{\eta_{i;\,\nu}}{2}\nonumber
\left[e^{i a \mu_{f,I}\delta_{\nu,4}}U^{(2)}_{i;\,\nu}\delta_{i,j-\hat{\nu}} \right. \nonumber\\
&-&\left. e^{-i a \mu_{f,I}\delta_{\nu,4}}U^{(2)\dagger}_{i-\hat\nu;\,\nu}\delta_{i,j+\hat\nu}  \right] \, .
\end{eqnarray}
where $U$ are the gauge link variables, $\mathcal{S}_{Y\!M}$ is the tree level
improved Symanzik gauge action~\cite{weisz,curci}, written in terms of
$W^{n\!\times \! m}_{i;\,\mu\nu}$ (trace of the $n\times m$ loop
constructed from the gauge links along the directions $\mu, \nu$ departing from
the $i$ site).  Finally, the staggered Dirac operator
$(M^f_{\textnormal{st}})_{i,\,j}$ is built up in terms of the two times
stout-smeared~\cite{morning} links $U^{(2)}_{i;\,\nu}$, with an isotropic
smearing parameter $\rho = 0.15$.  Stout smearing improvement is used in order
to reduce taste symmetry violations (see Ref.~\cite{bazavov} for a comparison
among different improved staggered discretizations); the rooting procedure is
exploited, as usual, to remove the residual fourth degeneracy of the staggered
lattice Dirac operator (see, e.g., Ref.~\cite{rooting} for a discussion on
possible related systematics).

The temperature of the system is given by $T = 1/(N_t a)$, where $a$ is the
lattice spacings and $N_t$ is the number of lattice sites in the temporal
direction, along which we take thermal boundary conditions
(periodic/antiperiodic for boson/fermion fields).  At fixed $N_t$, $T$ is
changed by varying the value of the bare coupling constant $\beta$.  The bare
quark masses $m_s$ and $m_{l}$ are rescaled accordingly, in order to move on a
line of constant physics, with $m_{\pi}\simeq 135\,\mathrm{MeV}$ and
$m_s/m_{l}=28.15$. This line is determined by a spline interpolation of the
values reported in Refs.~\cite{tcwup1,befjkkrs} (see also Ref.~\cite{crow}).
Four different sets of lattice spacings, corresponding to $N_t = 6,8,10,12$,
have been explored, in order to extrapolate our results to the continuum limit.

\subsection{Setup of chemical potentials}

In Eq.~(\ref{partfunc}), we have introduced an imaginary 
chemical potential
$\mu_f=i\mu_{f,I},\ \mu_{f,I}\in\mathbb{R}$,
with $f=u,d,s$, coupled to the number operator of each quark flavor.
They are related to the chemical potentials coupled to conserved
charges (baryon number $B$, electric charge $Q$ and strangeness $S$)
by the following relations
\begin{eqnarray}\label{defchem1}
\mu_u &=& \mu_B/3 + 2 \mu_Q/3 \nonumber \\
\mu_d &=& \mu_B/3 - \mu_Q/3 \\
\mu_s &=& \mu_B/3 - \mu_Q/3 -\mu_S \, .\nonumber
\end{eqnarray}
The purpose of our study is to determine the dependence of the pseudocritical
temperature $T_c$ on the baryon chemical potential (which
is given by $\mu_B = \mu_u + 2 \mu_d$),
in a setup of chemical potentials which is as close as possible to the
thermal equilibrium conditions created in heavy ion collisions.
We thus have to require
to $S=0$ and $Q=rB$, where $r$ is the number of protons divided by the number
of nucleons of the colliding ions, $r \equiv Z/A \approx 0.4$ typically.

These requirements can be translated into relations between $\mu_B$, $\mu_S$
and $\mu_Q$, which at the lowest order in $\mu_B$ read $\mu_Q\simeq q_1 (T)
\mu_B$ and $\mu_S\simeq s_1 (T) \mu_B$, the coefficients $q_1(T)$ and $s_1(T)$
being related to derivatives of the free energy
density~\cite{strangeness,strangeness2}. Let us consider as an example the
strangeness neutrality condition: in a gas of non-interacting fermions it would
imply $\mu_s = 0$ but in QCD, due to interactions, the mixed derivatives of the free
energy density with respect to $\mu_s$ and $\mu_u, \mu_d$ are non-vanishing, so
that one needs a non-zero $\mu_s$ to ensure $S=0$.  Present lattice
investigations~\cite{strangeness,strangeness2} show that, for $T\sim
155\,\mathrm{MeV}$, the constraints on charge and strangeness imply $s_1\simeq
0.25$ and $q_1\simeq -0.025$. With a precision of a few percent, 
around the transition at vanishing density, we thus have 
$\mu_l\equiv\mu_u=\mu_d$, $\mu_l\simeq \mu_B/3$ and $\mu_s \simeq \mu_l/4$.

Our determination of the curvature $\kappa$ has been obtained setting $\mu_s =
0$, which is close to the conditions described above. To quantify the impact of
$\mu_s$, as in Ref.~\cite{crow}, we have considered also the case $\mu_s =
\mu_l$, in order to obtain an estimate about the effect of a non-zero $\mu_s$
in a range which covers the equilibrium conditions created in heavy ion
collisions.

\subsection{Physical observables, renormalization and the 
determination of $T_c$}

In the absence of a true phase transition, the determination of the 
pseudo-critical line may depend on the physical observable chosen 
to locate it. On the other hand, chiral symmetry restoration
is the leading phenomenon around $T_c$, with
the light chiral condensate becoming
an exact order parameter in limit of zero light quark masses.
Therefore in the following $T_c (\mu_B)$ will be determined
by monitoring the chiral properties of the system.
The chiral condensate of the flavor $f$ is defined as
\begin{equation}
\langle\bar{\psi}\psi\rangle_f=\frac{T}{V}\frac{\partial \log Z}{\partial m_f}\ ,
\end{equation} 
where $V$ is the spatial volume. In our simulations the two light quarks 
are degenerate, $m_l\equiv m_u=m_d$, and it is convenient to introduce the
light quark condensate:
\begin{equation}
\langle\bar\psi\psi\rangle_l=\frac{T}{V}\frac{\partial \log Z}{\partial m_l}=
\langle\bar{u}u\rangle+\langle\bar{d}d\rangle\ ;
\end{equation} 
$\langle\bar\psi\psi\rangle_l$ is affected by both additive and
multiplicative renormalizations. We consider two different
renormalization prescriptions, in order to determine
whether any systematic effect related to this choice
affects the determination of $\kappa$. The first one~\cite{Cheng:2007jq}
is
\begin{equation} \label{rencond}
\langle\bar{\psi}\psi\rangle^r_{(1)}(T)\equiv\frac{\left[
\langle \bar{\psi}\psi\rangle_l -\frac{\ds 2m_{l}}{\ds m_s}\langle \bar{s}s\rangle\right](T)}{
\left[\langle \bar{\psi}\psi\rangle_l-\frac{\ds 2m_{l}}{\ds m_s}\langle \bar{s}s\rangle\right](T=0)}\ ,
\end{equation} 
where $m_s$ is the bare strange quark mass;
in this way the leading mass dependent contribution is 
subtracted\footnote{This prescription subtracts both divergent
and finite terms which are linear in the mass, thus permitting
to isolate contributions to the quark condensate directly related to 
spontaneous chiral
symmetry breaking. However, 
possibile additive logarithmic divergences could still be present.},
while one takes care of the multiplicative renormalization by dividing 
by the same quantity at $T = 0$.
As an alternative, we consider 
the following prescription~\cite{Endrodi2011}
\begin{equation}\label{eq:ren_pres_wupp}
\langle \bar{\psi}\psi\rangle^r_{(2)}=\frac{m_{l}}{m_{\pi}^4}\left(\langle\bar{\psi}\psi\rangle_{l}
-\langle\bar{\psi}\psi\rangle_{l}(T=0)\right)\, .
\end{equation}
In this case the zero $T$ subtraction eliminates additive divergences while
multiplication by the bare quark mass $m_l$ takes care of multiplicative
ones.

The behavior of both condensates will be monitored to locate $T_c$. In 
particular, since in the presence of a true phase transition the slope
of the condensate as a function of $T$ diverges at $T_c$, we will look
for the point of maximum slope, i.e.~the inflection point 
(a detailed comparison with other prescriptions has been reported 
in Ref.~\cite{crow}). 

A much better probe is provided by the chiral susceptibility
$\chi_{\bar\psi\psi}$, which is itself divergent at $T_c$ in the presence of a
true transition: in this case the introduction of relevant parameters (finite
mass or finite volume) smooths the divergence, however looking for the maximum
of $\chi_{\bar\psi\psi}$ remains a well defined and univoque prescription for
locating the pseudo-critical temperature $T_c$.  On the lattice, the light
chiral susceptibility is given by ($M_l$ is the Dirac operator corresponding
to a single light flavor)
\begin{eqnarray}
\label{susc}
\chi_{\bar\psi\psi}
&=& \frac{\partial\langle\bar\psi\psi\rangle_l}{\partial m_l}
=\chi_{\bar\psi\psi}^{disc}+\chi_{\bar\psi\psi}^{conn}\\
\label{sconn}
\chi_{\bar\psi\psi}^{disc}&\equiv&\frac{T}{V}\left(\frac{N_l}{4}\right)^2
\left[\langle (\tr M_l^{-1})^2\rangle-\langle \tr M_l^{-1}\rangle^2 \right] \\
\label{conn}
\chi_{\bar\psi\psi}^{conn}&\equiv&
-\frac{T}{V}\frac{N_l}{4}\langle \tr M_l^{-2}\rangle\,.
\end{eqnarray}
where $N_l = 2$ is the number of degenerate light quarks.
The renormalization is performed by first
subtracting the $T=0$ contribution, to remove the additive renormalization,
then multiplying the result by the square of the bare light quark mass, to
cancel the multiplicative one~\cite{tcwup1}:
\begin{equation} \label{rensusc}
\chi_{\bar{\psi}\psi}^r=m_{l}^2\left[ \chi_{\bar{\psi}\psi}(T)-\chi_{\bar{\psi}\psi}(T=0)\right]\,.
\end{equation}

\subsection{Analytic continuation from imaginary chemical potentials}

The physical observables relevant to our study will be monitored
as a function of $T$ for fixed values of the 
dimensionless ratio $\theta_{l} = {\rm Im} (\mu_{l})/T$.
In this way we shall be able to locate $T_c$ for a set of
values of $\theta_l$, so as to determine the dependence 
${T_c(\theta_{l})}$ to the leading order
\begin{equation}
\frac{T_c(\theta_{l})}{T_c(0)} = 1 + R\, \theta_{l}^2 + O(\theta_{l}^4) \, ,
\label{quadratic4}
\end{equation}
where we have assumed $T_c (\theta_l)$ to be an analytic function
of $\theta_l$, at least for small values of it. This assumption 
is consistent with numerical data and is at the basis 
of the method of analytic continuation. Comparing 
with Eq.~(\ref{corcur}) one has, at the leading order in $\mu_B^2$,
$\kappa = R/9$.

\section{Numerical Results}
\label{sec_results}

We have performed simulations on lattices with $N_t = 8,10$ and $12$
and different choices of $T$ and of the chemical potentials; 
results will be combined with those already presented in Ref.~\cite{crow}
for $N_t = 6,8$ to perform the continuum extrapolation. To 
that purpose, we will consider only lattices 
with fixed aspect ratio $L_s/N_t = 4$: that guarantees the absence of 
significant finite size effects (see Ref.~\cite{crow} for a detailed
study about that).

Four different values of chemical potentials have been considered for $N_t =
10,12$, corresponding to $\mu_s = 0$ and ${\rm Im} (\mu_{l})/(\pi T) = 0, 0.20,
0.24$ and 0.275.  A larger set has been considered for $N_t = 8$, in which case
we performed simulations also at $\mu_s \neq 0$, in order to provide more
information about systematics related to the choice of $\mu_s/\mu_l$ and to the
truncation of the Taylor expansion in Eq.~(\ref{quadratic4}).

\begin{figure*}[t!]
\begin{tabular}{cc}
\includegraphics[width=0.92\columnwidth, clip]{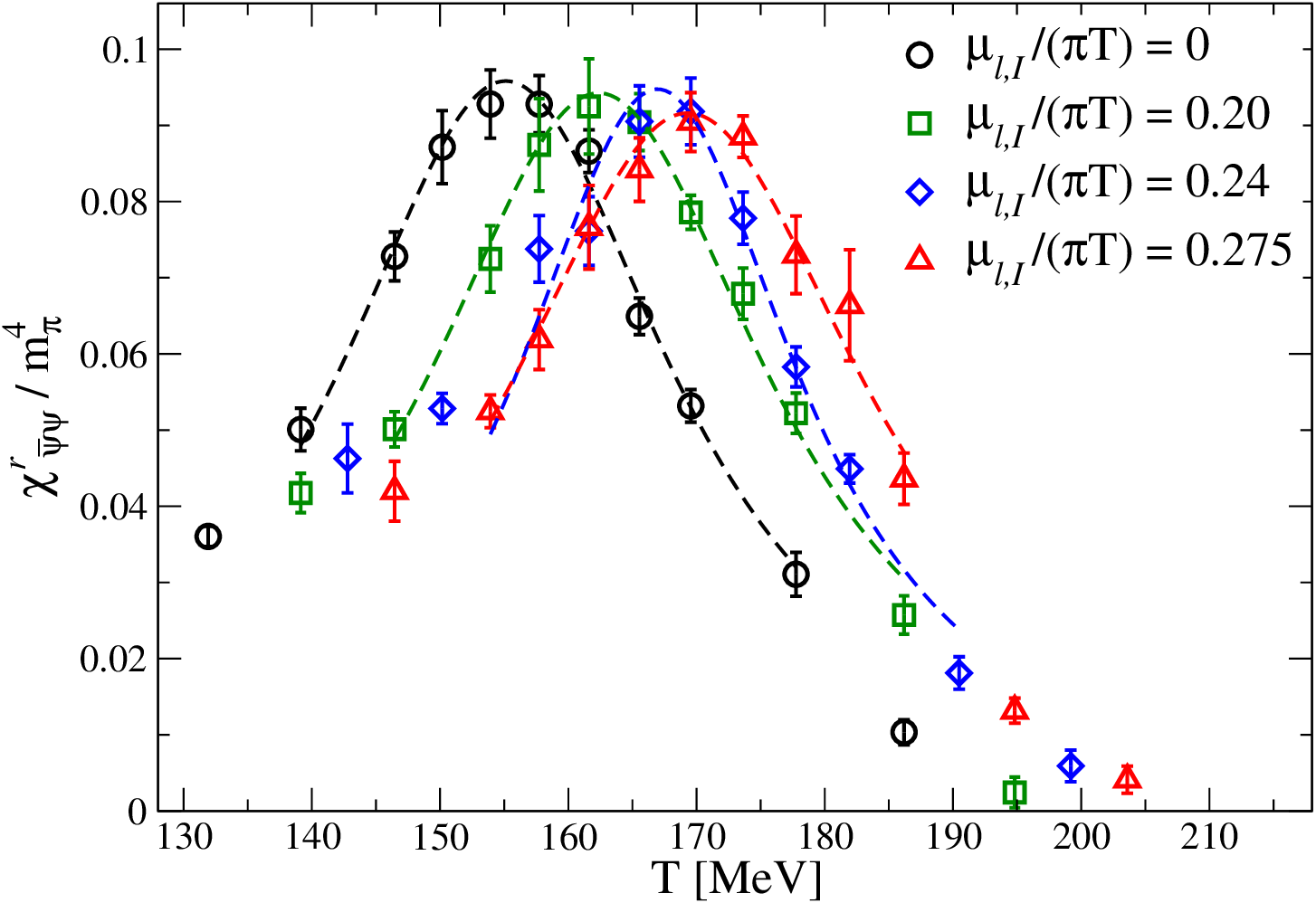} & \includegraphics[width=0.92\columnwidth, clip]{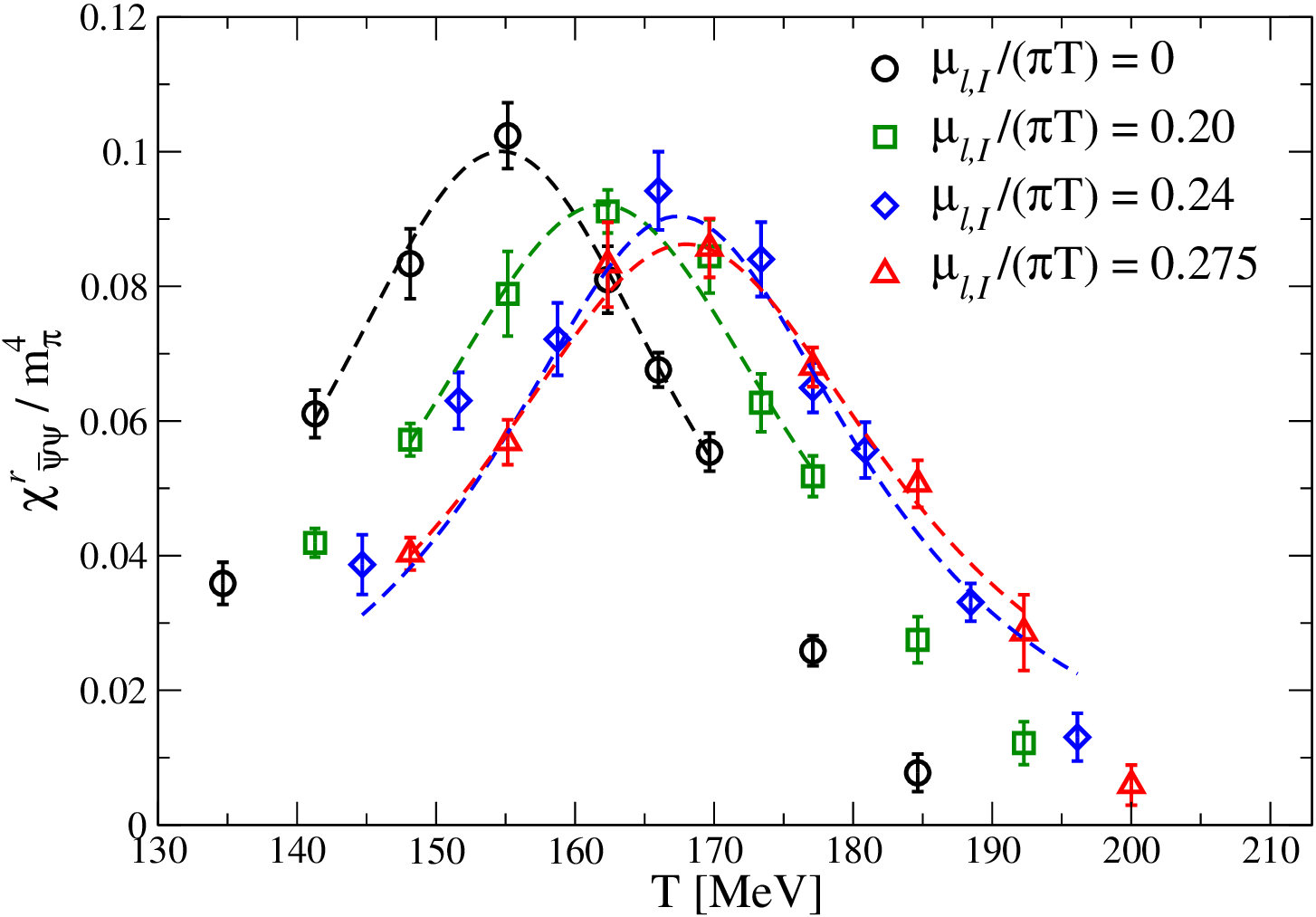} \\
\includegraphics[width=0.92\columnwidth, clip]{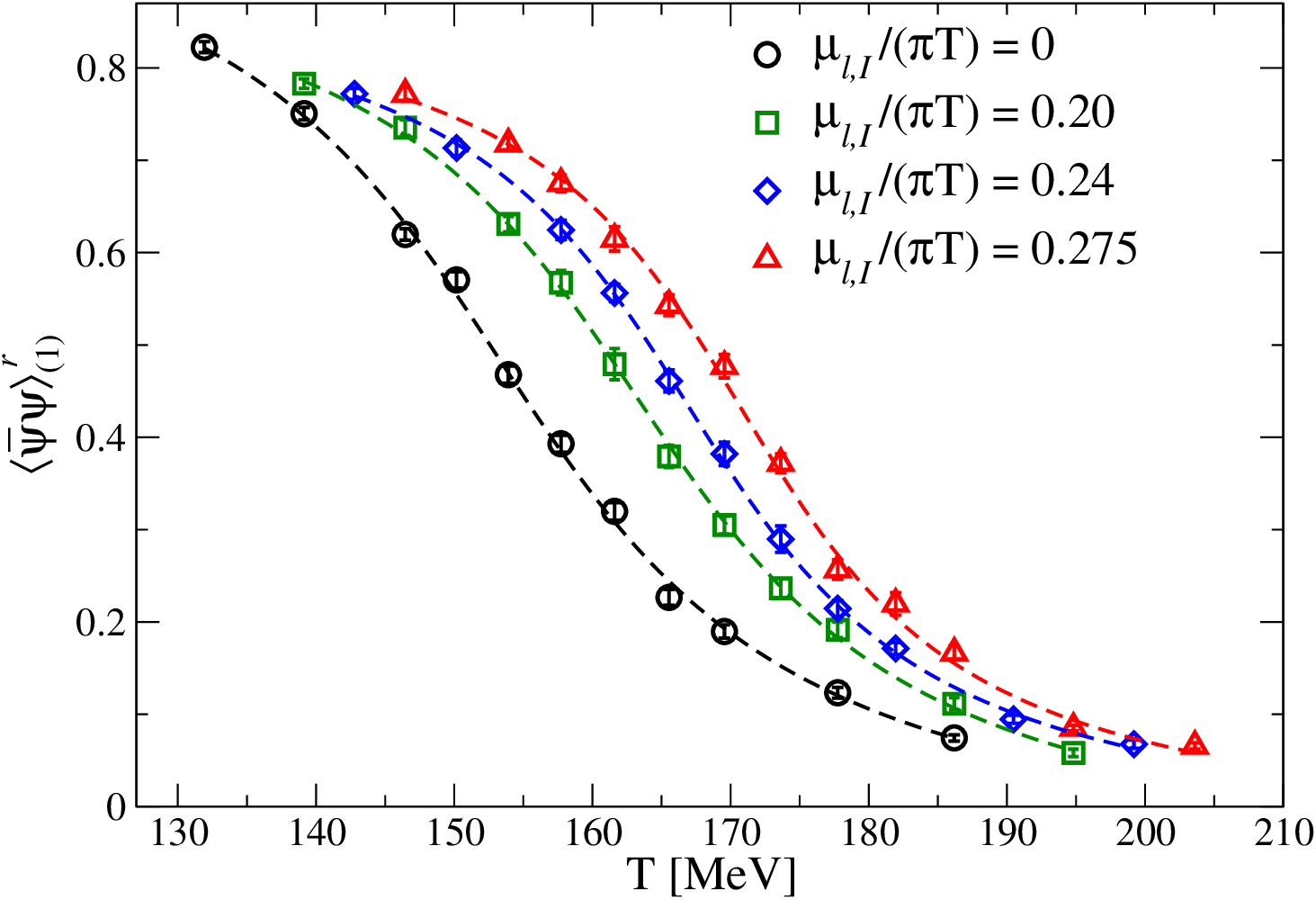} & \includegraphics[width=0.92\columnwidth, clip]{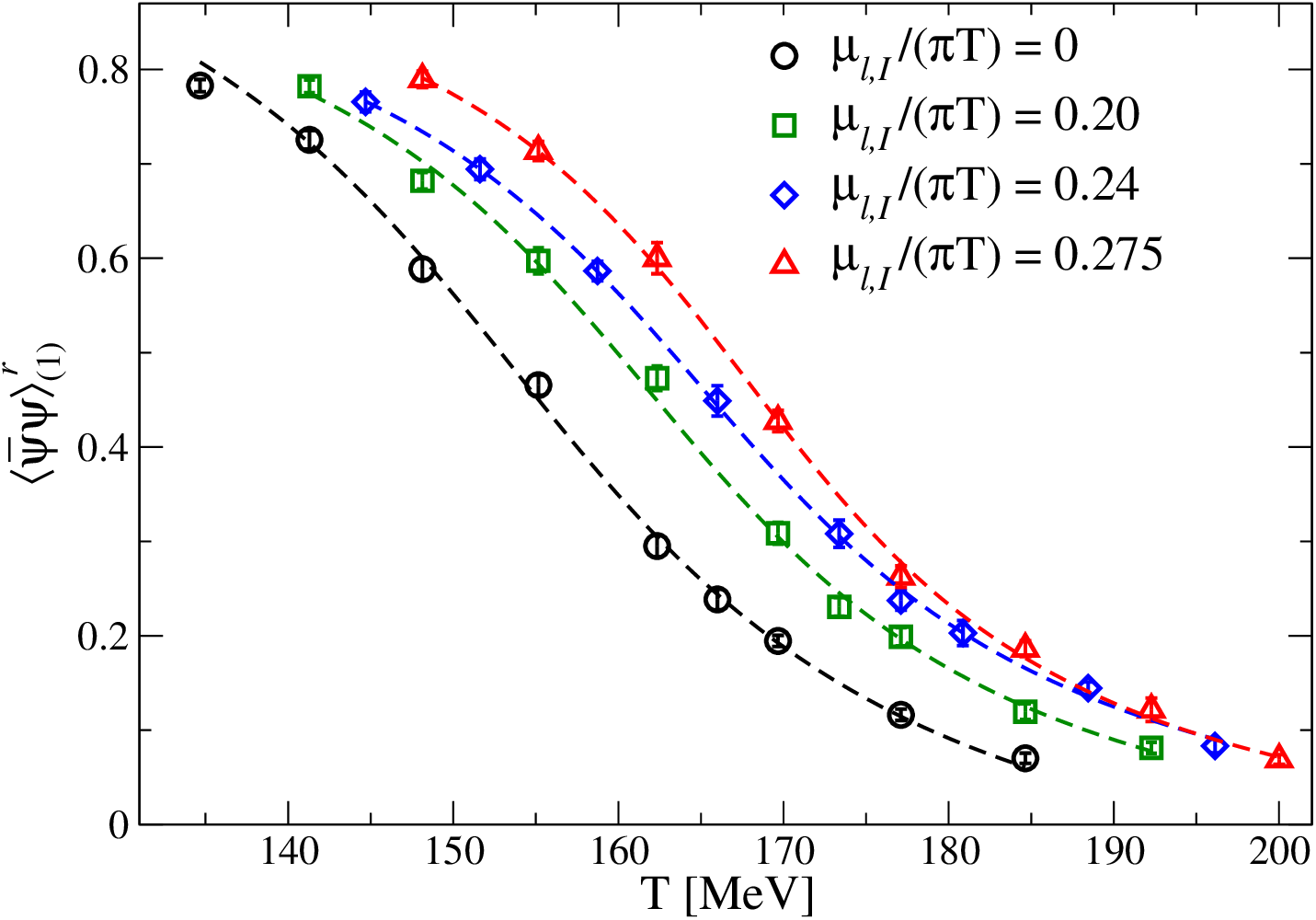} \\
\includegraphics[width=0.92\columnwidth, clip]{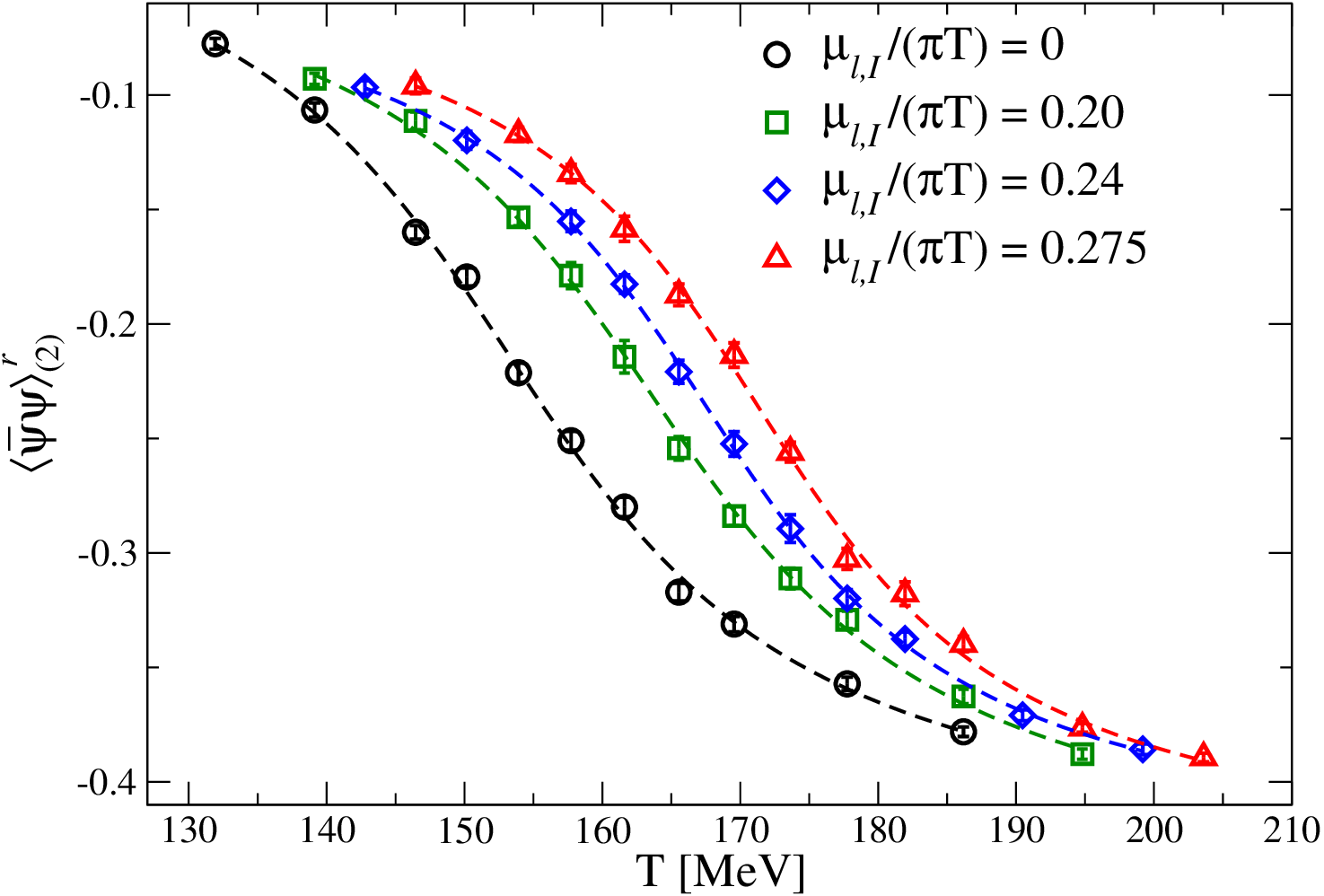} & \includegraphics[width=0.92\columnwidth, clip]{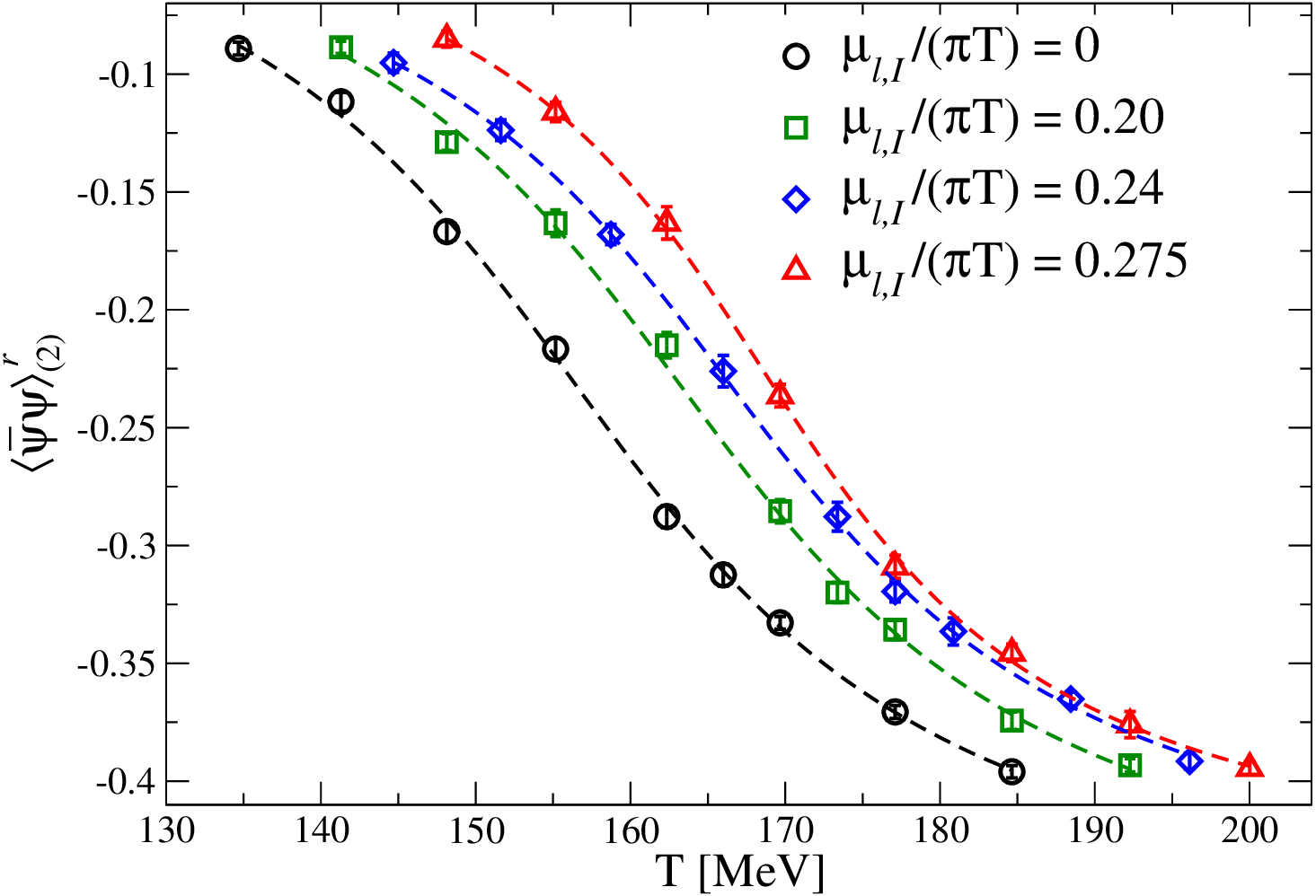} \\
\end{tabular}
\caption{Renormalized susceptibility and chiral condensates for the 
$40^3\times 10$ (left column) and $48^3\times 12$ lattices (right column).}
\label{cond404812}
\end{figure*}

\begin{table}[t!]
\centering
\begin{tabular}{ |c|c|c|c|c|c| }
\hline
\rule{-2mm}{4mm}
Lattice & $\frac{\mu_{l,I}}{\pi T}$ & $\frac{\mu_{s,I}}{\pi T}$ & 
$T_c(\bar{\psi}\psi_{(1)})$ &  
$T_c(\bar{\psi}\psi_{(2)})$ & 
$T_c(\chi^r)$ \\
& & & & & \\
\hline
$16^3\times 6$ & 0.00  & 0.00   &  148.2(3) &  148.4(4) &  150.7(4) \\ 
$16^3\times 6$ & 0.20  & 0.00   &  155.0(4) &  155.1(5) &  157.0(4)  \\ 
$16^3\times 6$ & 0.24  & 0.00   &  158.9(4) &  159.1(4) &  160.0(4)  \\ 
$16^3\times 6$ & 0.275 & 0.00   &  161.2(4) &  161.5(4) &  162.7(4)  \\ 
\hline                                                            
$24^3\times 6$ & 0.00  & 0.00   &  149.0(6) & 149.0(6) &  151.6(5) \\ 
$24^3\times 6$ & 0.24  & 0.00   &  160.8(7) & 160.7(5) &  162.0(5)  \\ 
$24^3\times 6$ & 0.275 & 0.00   &  164.1(4) & 164.3(5) &  165.9(4)  \\ 
\hline                                                               
$32^3\times 6$ & 0.00  & 0.00   &  149.1(7) &  149.4(4)  &  152.0(4) \\ 
$32^3\times 6$ & 0.24  & 0.00   &  160.2(3) &  160.4(2)  &  162.7(4)  \\ 
$32^3\times 6$ & 0.275 & 0.00   &  163.4(3) &  163.5(3)  &  165.5(4)  \\ 
\hline                                                               
$32^3\times 8$ & 0.00  & 0.00   &  154.2(4) &  154.5(4)  &  155.6(7)\\
$32^3\times 8$ & 0.10  & 0.00   &   155.4(7) &  155.2(8) &  157.2(7) \\
$32^3\times 8$ & 0.15  & 0.00   &   159.5(9) & 158.9(9) &  160.2(5) \\
$32^3\times 8$ & 0.20  & 0.00   &  162.9(8) &  163.0(6) &  163.0(6)  \\
$32^3\times 8$ & 0.24  & 0.00   &  165.0(5) &  164.8(5) &  165.8(8)  \\
$32^3\times 8$ & 0.275 & 0.00   &  169.5(9) &  168.6(7) &  169.8(7)  \\
$32^3\times 8$ & 0.30  & 0.00   &  172.4(9) &  171.8(9)&  172.8(8) \\
$32^3\times 8$ & 0.10  & 0.10   &   157.1(8) &  157.0(8) &  158.5(7)\\
$32^3\times 8$ & 0.15  & 0.15   &   159.2(9) &  158.8(8) &  160.1(8) \\
$32^3\times 8$ & 0.20  & 0.20   &  163.9(6) &  163.7(6) &  165.3(9)  \\
$32^3\times 8$ & 0.24  & 0.24   &  169.4(7) &  168.6(6) &  169.6(7)  \\
$32^3\times 8$ & 0.275 & 0.275  &  175.4(6) &  174.4(7) &  177.0(8)  \\
\hline                                                            
$40^3\times 10$ & 0.00  & 0.00   & 154.5(1.5)  & 154.3(1.5) &  155.1(7)  \\ 
$40^3\times 10$ & 0.20  & 0.00   &  163.0(7) &  163.0(8) &   162.5(7)  \\ 
$40^3\times 10$ & 0.24  & 0.00   &  166.8(8) & 167.1(7) &  166.2(1.0)  \\ 
$40^3\times 10$ & 0.275 & 0.00   &  170.8(8) & 171.2(8) &  169.6(8)  \\ 
\hline                                                            
$48^3\times 12$ & 0.00  & 0.00   & 154.5(1.0)  & 155.5(1.3) &  154.7(7)  \\ 
$48^3\times 12$ & 0.20  & 0.00   &  163.2(1.2) & 165.0(1.5) &   161.9(7) \\ 
$48^3\times 12$ & 0.24  & 0.00   & 165.2(1.1)  & 166.2(1.0) &  166.2(1.0)  \\ 
$48^3\times 12$ & 0.275 & 0.00   &  167.8(1.2) & 168.7(9) &   167.9(9) \\ 
\hline
\end{tabular}
\caption{Critical values of $T$ obtained from the renormalized chiral susceptibility
and from the renormalized chiral condensates. Errors do not take
into account the uncertainty on the physical scale, which
is of the order of 2-3\,\%~\cite{tcwup1,befjkkrs}.} \label{tab:tc_all}
\end{table}

For each setup of chemical potentials we have explored $\mathcal{O}(10)$
different temperatures around $T_{c}(\theta_l)$.  The Rational Hybrid
Monte-Carlo algorithm~\cite{rhmc1, rhmc2, rhmc3} has been used to sample gauge
configurations according to Eq.~(\ref{partfunc}), each single run consisting of
2-5 K trajectories of unit length in molecular dynamics time, with higher
statistics around the transition. 

Traces appearing in the definition of chiral quantities (see, e.g.,
Eqs.~(\ref{sconn}) and (\ref{conn})) have been computed by noisy estimators
{at the end of each molecular dynamics trajectory,  
using $8$ random vectors for each flavor. Such a choice has appeared,
after some preliminary tests, 
as a reasonable compromise to balance the effort spent in the 
stochastic estimators and in the gauge configuration production,
i.e. in order to optimize the statistical error obtained 
for a given computational effort.}
  A jackknife analysis has been
exploited to determine the statistical errors.

To perform the renormalization described in Sec.~\ref{sec2}, one needs to
compute observables also at $T=0$ and at the same values of the bare
parameters, i.e. at the same ultraviolet (UV) cutoff. For that reason we have
performed simulations on lattices as large as $48^4$: details are reported in
Appendix~\ref{app:t0}.

In order to determine the inflection point of the renormalized chiral
condensate, we have performed a best fit to our data according to
\begin{equation}
\label{arctan}
\langle \bar{\psi}\psi\rangle^{r} (T) = A_1  + B_1 \arctan\left(C_1 (T-T_{c})\right)\, ,
\end{equation}
which involves the independent parameters
$A_1$, $B_1$, $C_1$ and $T_c$. Instead, for the determination
of the peak of the renormalized susceptibility, we have
performed a best fit according to a Lorentzian function
\begin{equation}\label{lorentz}
\chi^r_{\bar{\psi}\psi} = \frac{A_2}{B_2^2 + (T-T_{c})^2}\, .
\end{equation}
Both functions are found to well describe respective data points around $T_c$.
In both cases, statistical errors on the fitted parameters have been estimated
by means of a bootstrap analysis, while systematic uncertainties have been
estimated either by varying the range of fitted points around $T_c$ or by
choosing an alternative fitting function (e.g., a hyperbolic tangent for the
condensate or a parabola for its susceptibility).  Statistical and
systematic\footnote{We do not report the systematic error on the determination
of the physical scale, which is of the order of 2-3\,\%~\cite{tcwup1,befjkkrs}
and, being related to an overall scale determination, does not affect the ratio
of pseudocritical temperatures entering the determination of $\kappa$, see
Eqs.~(\ref{corcur}) and (\ref{quadratic4}).} errors are both included in the
collection of determinations of $T_c$ for the various combinations of lattice
sizes and chemical potentials in Table~\ref{tab:tc_all}, which includes, for
completeness, also results presented in Ref.~\cite{crow}.

In Fig.~\ref{cond404812} we report results obtained
for $\chi^r_{\bar{\psi}\psi}$, $\langle \bar{\psi}\psi\rangle^{r}_{(1)}$ and 
$\langle \bar{\psi}\psi\rangle^{r}_{(2)}$
on the $40^3 \times 10$ and $48^3 \times 12$ lattice, together
with some best fits according to Eqs.~(\ref{arctan}) and (\ref{lorentz}).
In the following we will perform the continuum limit using two different
methods, in order to check for systematics effects.

\subsection{Continuum limit for $\mu_s = 0$ - First method}

\begin{figure}[t!]
\includegraphics[width=0.92\columnwidth, clip]{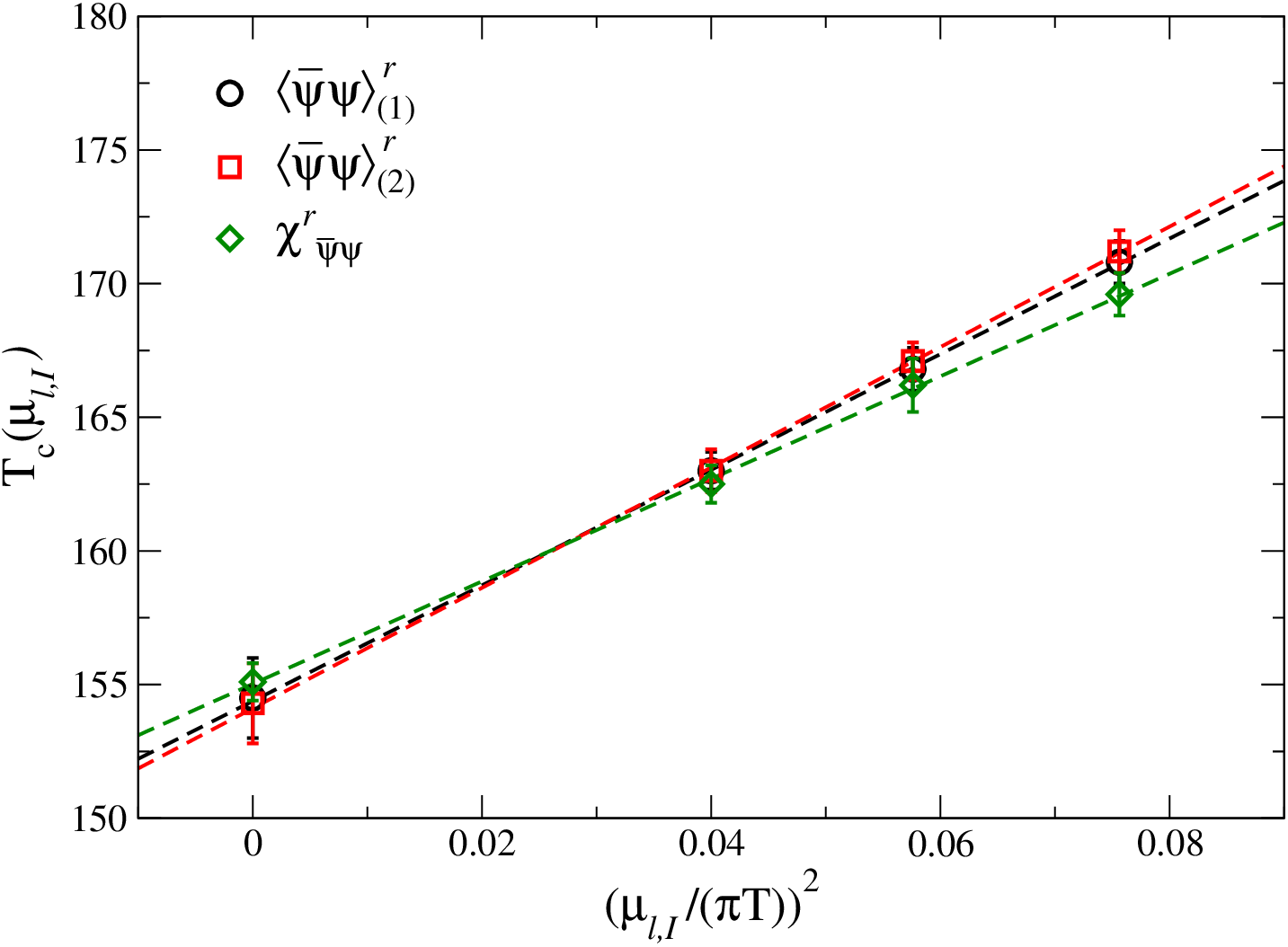}
\includegraphics[width=0.92\columnwidth, clip]{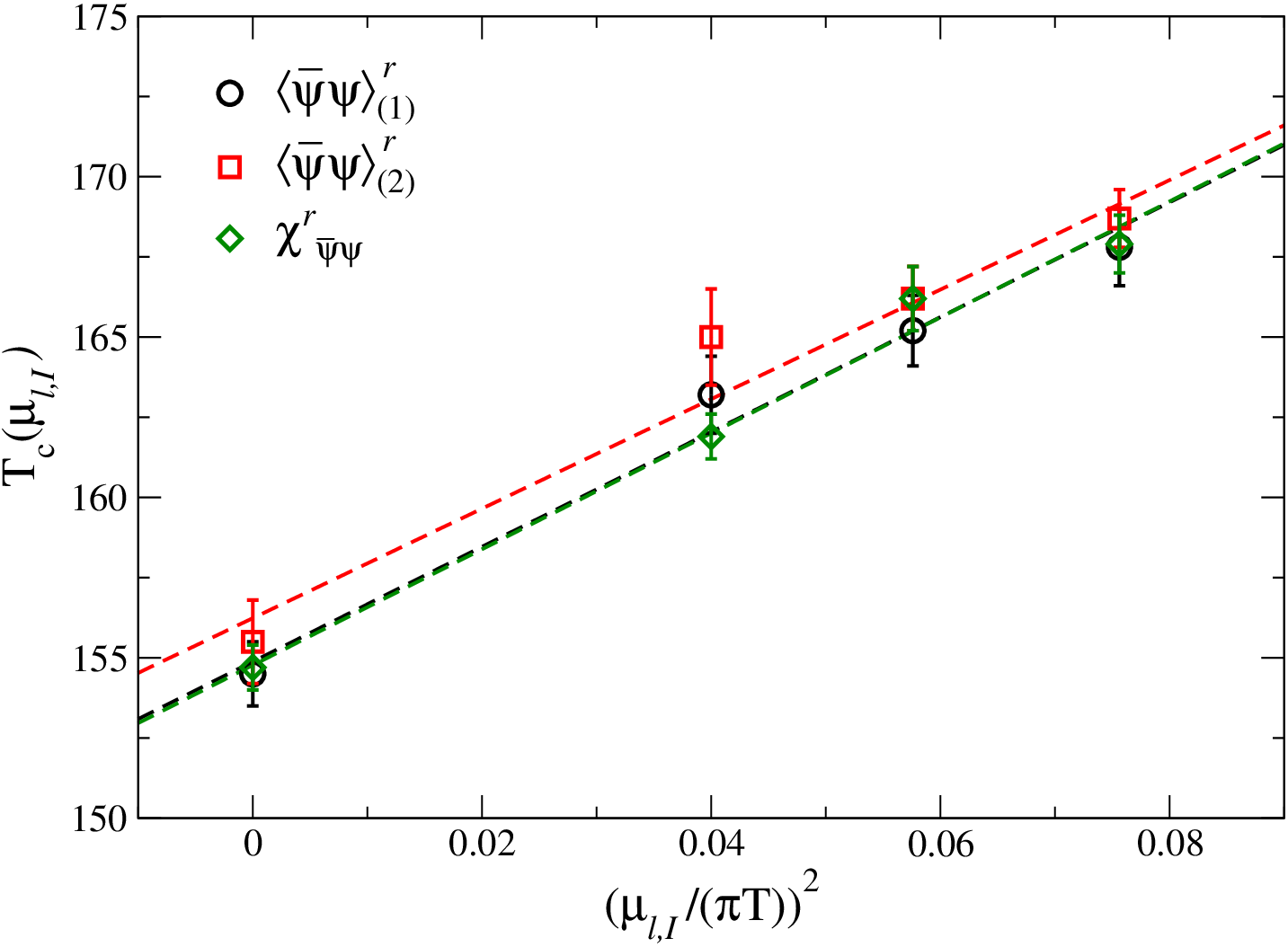}
\caption{Critical lines for the $40^3\times 10$ lattice (top) and
for the $48^3\times 12$ one (bottom).}
\label{fitlines}
\end{figure}


\begin{table}[b!]
\centering
\begin{tabular}{|c|c|c|c|c|c|}
\hline
Lattice & $\kappa (\bar{\psi}\psi_{(1)}) $  
& $\kappa (\bar{\psi}\psi_{(2)}) $  
& $\kappa (\chi^r) $  
\\
\hline
$24^3\times 6$ & 0.0150(7)  & 0.00152(7) & 0.0140(7)  \\
$32^3\times 8$ & 0.0142(7)  &  0.0135(7) & 0.0134(9)  \\
$40^3\times 10$ & 0.0157(17) & 0.0164(16) & 0.0139(10) \\ 
$48^3\times 12$ & 0.0130(15)  & 0.0123(17) &  0.0131(11)\\
\hline
\end{tabular}
\caption{Curvatures obtained at fixed $N_t$ from different observables.}
\label{tab:fit_kappa}
\end{table}

In order to extract the curvature of the critical line, we have performed a
best fit to the values of $T_c(\mu_{l,I})$, obtained for each lattice
size and setup of chemical potentials, according to the function
\begin{equation}
T_c(\mu_{l,I}) = T_c(0)\, \left( 1 + 
9 \kappa \left( \frac{\mu_{l,I}}{T_c(\mu_{l,I})} \right)^2 + 
 O(\mu_{l,I}^4) \right)\, .
\label{fitfun}
\end{equation}
For all sets of chemical potentials explored for 
$\mu_s = 0$, the inclusion of quartic corrections has not been necessary:
a more detailed discussion about the stability of the fit
as the range of chemical potentials is changed is reported in 
Sec.~\ref{mu_mu}.

In Fig.~\ref{fitlines} we report an example of such quadratic fits
to the critical temperatures obtained for $N_t = 10,12$ and
for the various explored observables.
A complete collection of results, including also those
already presented in Ref.~\cite{crow}, is reported in 
Table~\ref{tab:fit_kappa}.

\begin{figure}[t!]
\includegraphics[width=0.92\columnwidth, clip]{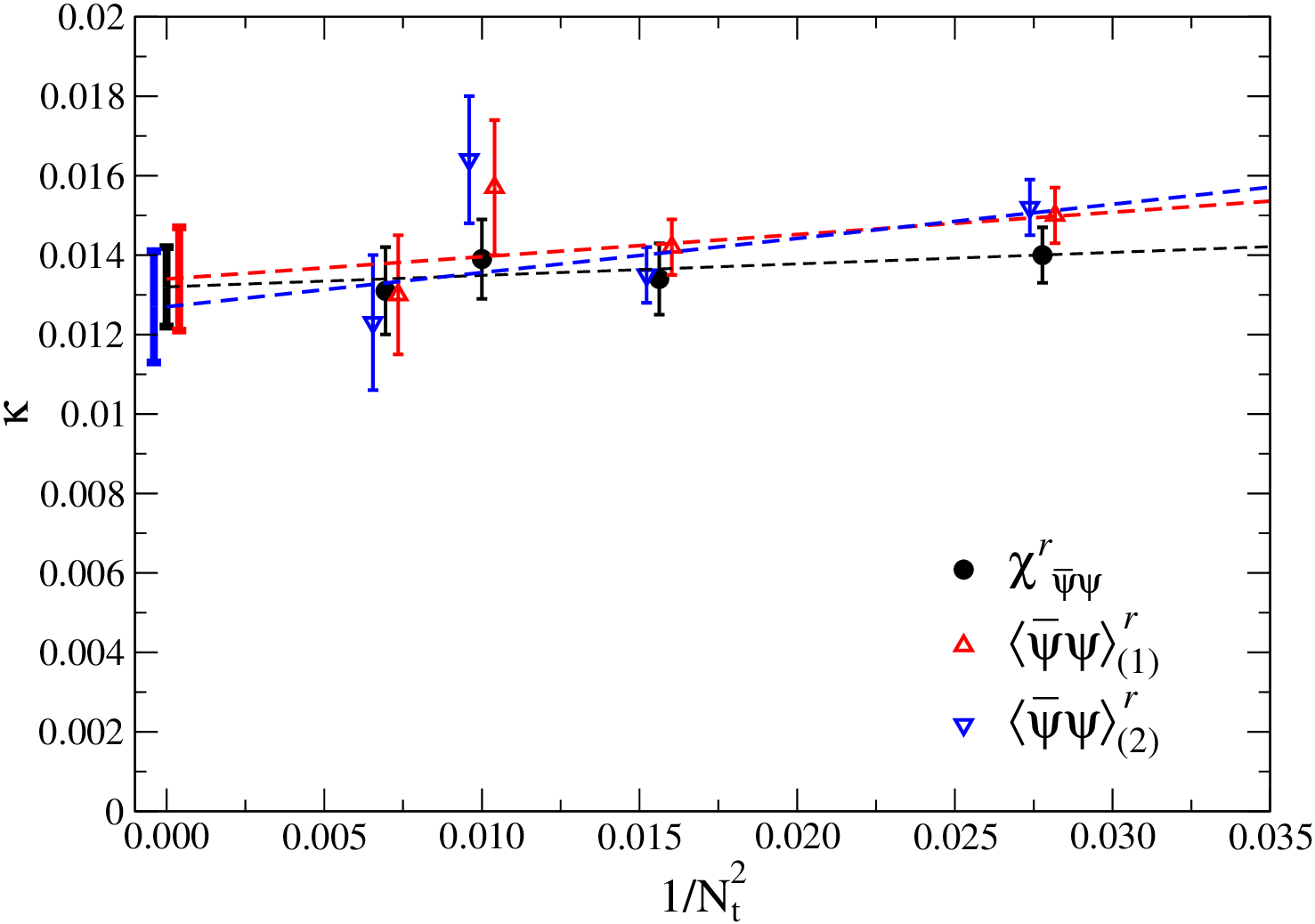}
\caption{Continuum limit of the curvatures extracted at fixed $N_t$ (data have
been slightly shifted in the horizontal direction to improve readability).}
\label{cont_1}
\end{figure}

In a range of temperatures around $T_c$, the UV cutoff 
$a^{-1}$ is approximately proportional to $N_t$. Therefore,
assuming corrections proportional to $a^2$, we extracted,
from the curvatures obtained for different values of $N_t$, 
continuum extrapolated results according to the 
ansatz
\beq
\kappa (N_t) = \kappa_{cont} + {\rm const.}/N_t^2   \, .
\eeq 
Results are shown in Fig.~\ref{cont_1}, where we also 
report the extrapolated continuum values, which are
$\kappa_{cont} (\langle\bar{\psi}\psi\rangle^r_{(1)}) = 0.0134(13)$,
$\kappa_{cont} (\langle\bar{\psi}\psi\rangle^r_{(2)}) = 0.0127(14)$
and
$\kappa_{cont} (\chi_{\bar\psi\psi}^r) = 0.0132(10)$.

\subsection{Continuum limit for $\mu_s = 0$ - Second method}

Results of the previous section show that the continuum extrapolation of
$\kappa$ is quite smooth, with a good agreement between the results obtained
with different observables and different renormalization prescriptions. This is
also consistent with the preliminary evidence reported in Ref.~\cite{crow}.

Nevertheless, it is useful to explore different ways of performing the
continuum limit, in order to check for the overall consistency of the
procedure.  In the previous section we first determined the value of $\kappa$
at each single value of $N_t$, then extrapolated these results to $N_t \to
\infty$ to obtain $\kappa_{cont}$.  A different procedure is to first
extrapolate the critical temperatures to $N_t \to \infty$ (for fixed values of
the dimensionless ratio $\mu_{l,I}/T$) and then to extract the value of
$\kappa_{cont}$ by using the continuum extrapolated critical temperatures.

To implement the second procedure we have performed, separately for each
$\mu_{l,I}/T$, a best fit to the values obtained for the renormalized
condensates and for the renormalized chiral susceptibility on different values
of $N_t$, according to modified versions of Eqs.~(\ref{arctan}) and
(\ref{lorentz}).  Since the cut-off dependence is more pronounced for such
quantities, we have excluded $N_t = 6$ data, thus using only $N_t =
8,10,12$.
 
\begin{figure}[t!]
\includegraphics[width=0.92\columnwidth, clip]{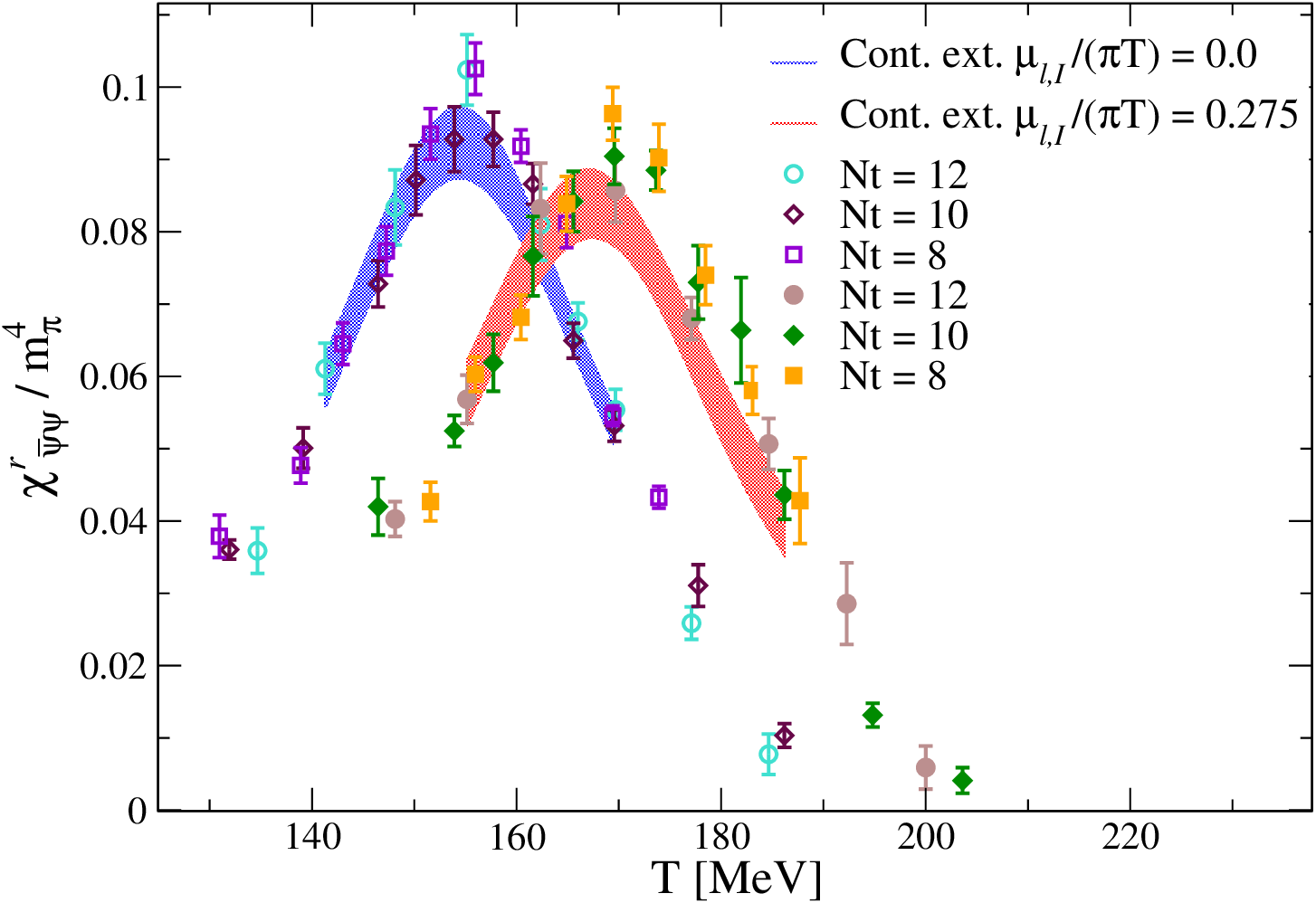}
\includegraphics[width=0.92\columnwidth, clip]{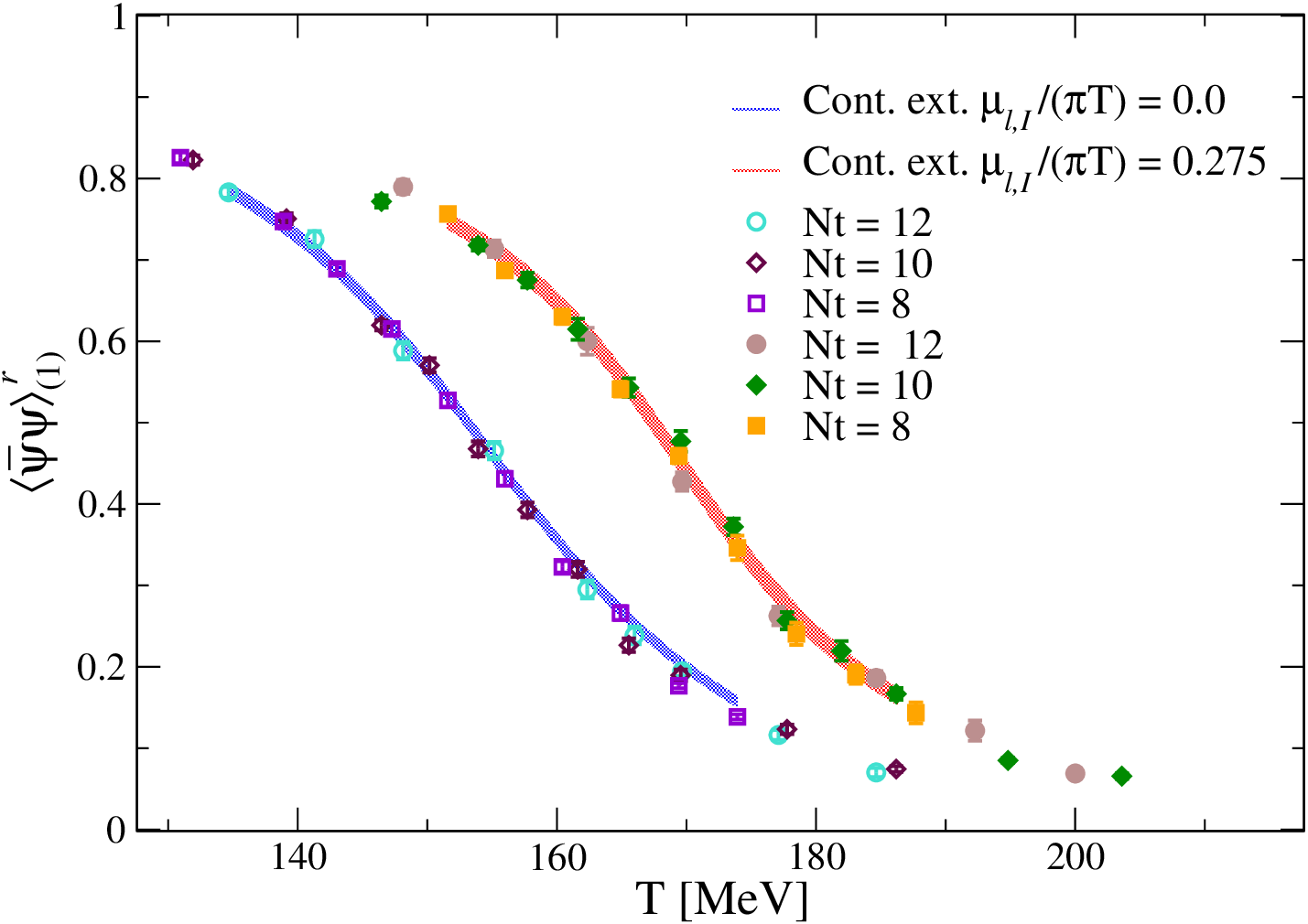}
\includegraphics[width=0.92\columnwidth, clip]{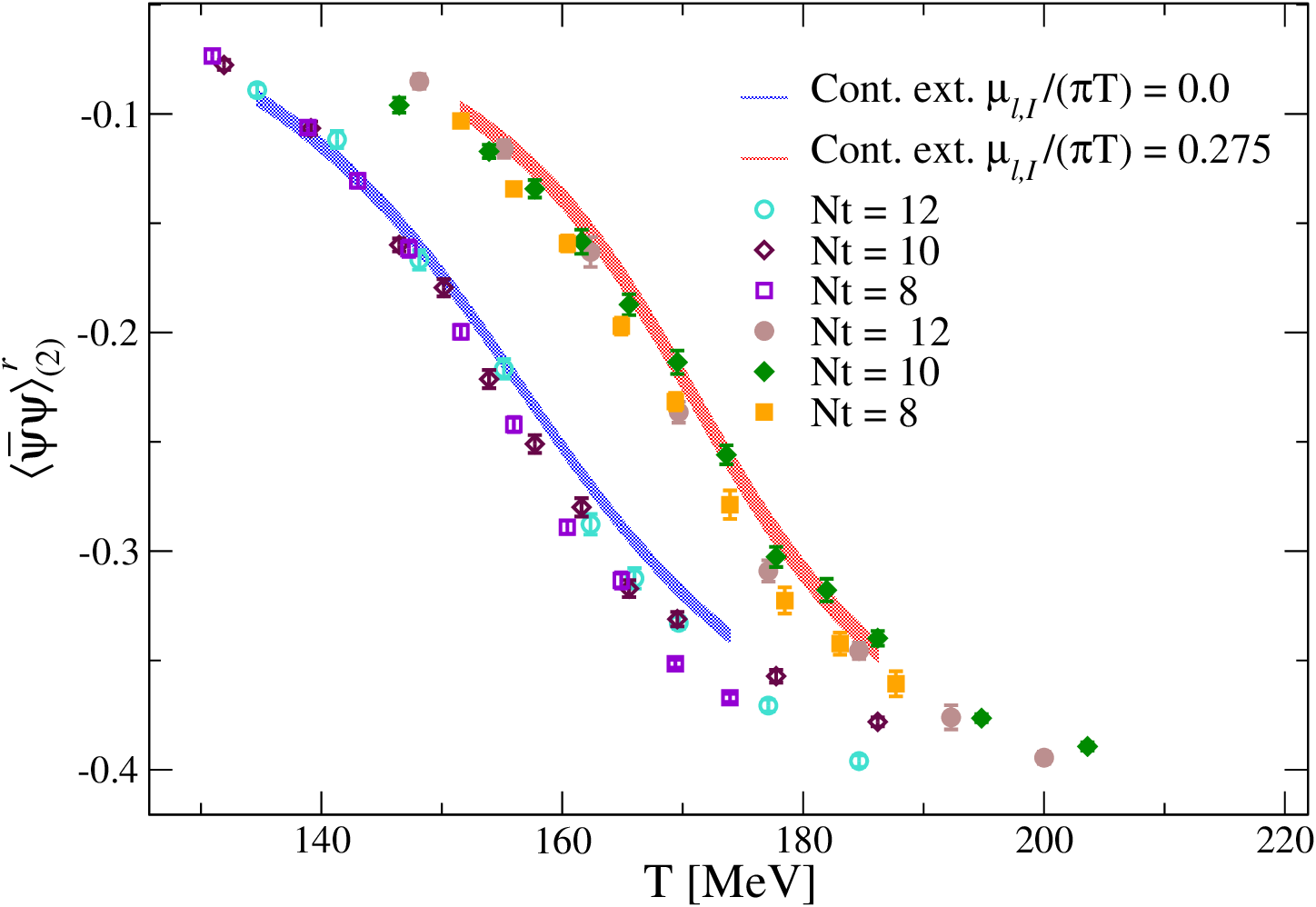}
\caption{Continuum limit for the renormalized susceptibility and the 
renormalized chiral condensates.}
\label{suscband}
\end{figure}

In detail, in the case of the renormalized susceptibility, each fit parameter
appearing in Eq.~(\ref{lorentz}) has been given an additional $N_t$ dependence,
for instance $T_c(N_t) = T_c (N_t = \infty) + {\rm const}/N_t^2$.  Results for
the extrapolated quantities are reported in the upper plot in
Fig.~\ref{suscband} where, for the sake of clarity, we report only the cases
$\mu_{l,I} = 0$ and $\mu_{l,I}/(\pi T) = 0.275$. In the case of the
renormalized condensates, instead, due to the larger number of parameters which
are present in Eq.~(\ref{arctan}), we could obtain fits which are stable
against the variation of the fitted range by adding an $N_t$-dependence to just
two parameters, in particular $T_c$ and $C_1$. Results are shown in the middle
and lower plot of Fig.~\ref{suscband}.

\begin{figure}[t!]
\includegraphics[width=0.92\columnwidth, clip]{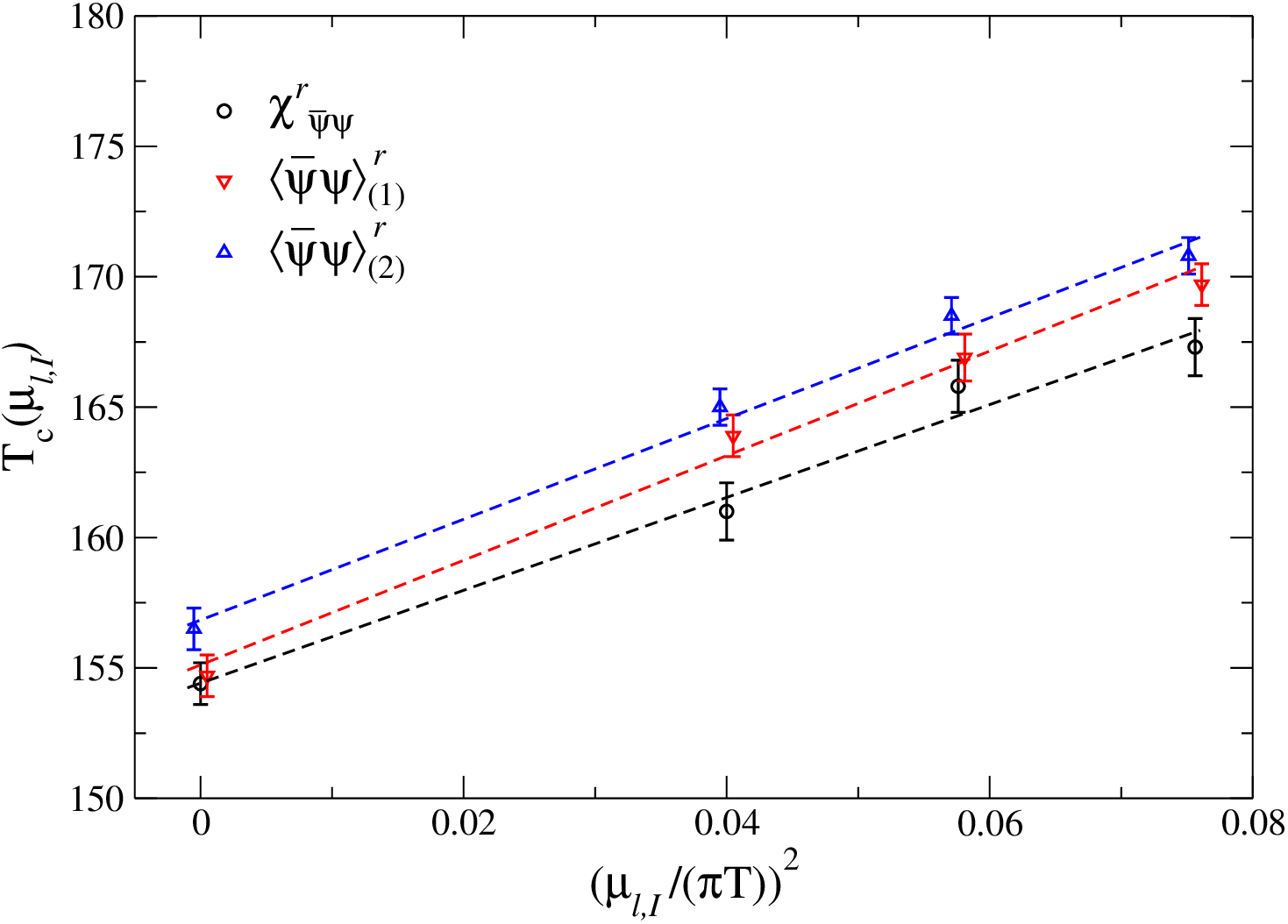}
\caption{Critical lines obtained by using the continuum extrapolated
renormalized chiral susceptibility and the continuum 
extrapolated chiral condensates.}
\label{tcont_fig}
\end{figure}

Such fits provide estimates for the continuum extrapolated pseudo-critical
temperatures, reported in Table~\ref{tab:cont_temp} and in
Fig.~\ref{tcont_fig}.  Such values coincide, within errors, with the continuum
pseudo-critical temperatures that one could obtain by directly fitting results
reported in Table~\ref{tab:tc_all}.  A best fit to the extrapolated
temperatures according to Eq.~(\ref{fitfun}), with only the quadratic term
included, provides $\kappa_{cont} (\langle\bar{\psi}\psi\rangle^r_{(1)}) =
0.0145(11)$, $\kappa_{cont} (\langle\bar{\psi}\psi\rangle^r_{(2)}) =
0.0138(10)$ and $\kappa_{cont} (\chi_{\bar\psi\psi}^r) = 0.0131(12)$, which are
consistent with those found previously.

\begin{table}[h]
\centering
\begin{tabular}{|c|c|c|c|c|c|}
\hline
$\mu_{l,I}/(\pi T) $ & $T_c (\bar{\psi}\psi_{(1)}) $  
& $T_c (\bar{\psi}\psi_{(2)}) $  
& $T_c (\chi^r) $  
\\
\hline
0.00  & 154.7(8) & 156.5(8) & 154.4(8) \\
0.20  & 163.9(8) & 165.0(7) & 161.0(1.1) \\
0.24  & 166.9(9) & 168.5(7) & 165.8(1.0) \\ 
0.275 & 169.7(8) & 170.8(7) & 167.3(1.1) \\
\hline
\end{tabular}
\caption{Continuum extrapolated critical temperatures for the various 
$\mu_{l, I}$ values.} 
\label{tab:cont_temp}
\end{table}

\subsection{Strength of the transition as a function of $\mu_B$}

The width and the height of the chiral susceptibility peak, which can be
obtained respectively from $B_2$ and $A_2/B_2^2$ in Eq.~(\ref{lorentz}), are
directly related to the strength of the chiral pseudo-transition. Therefore, we
have the possibility to monitor the dependence of such strength on the baryon
chemical potential and, having performed a continuum extrapolation for
$\chi_{\bar\psi\psi}^r$, we can do that directly on continuum extrapolated
quantities.

If a critical endpoint exists, along the pseudo-critical line, for relatively
small values of real $\mu_B$, we might expect a visible dependence of the
strength parameters also for small values of imaginary $\mu_B$.  The width and
the height would tend respectively to zero and infinity approaching, e.g., a
critical endpoint in the $Z_2$ universality class.

\begin{figure}[t!]
\includegraphics[width=0.92\columnwidth, clip]{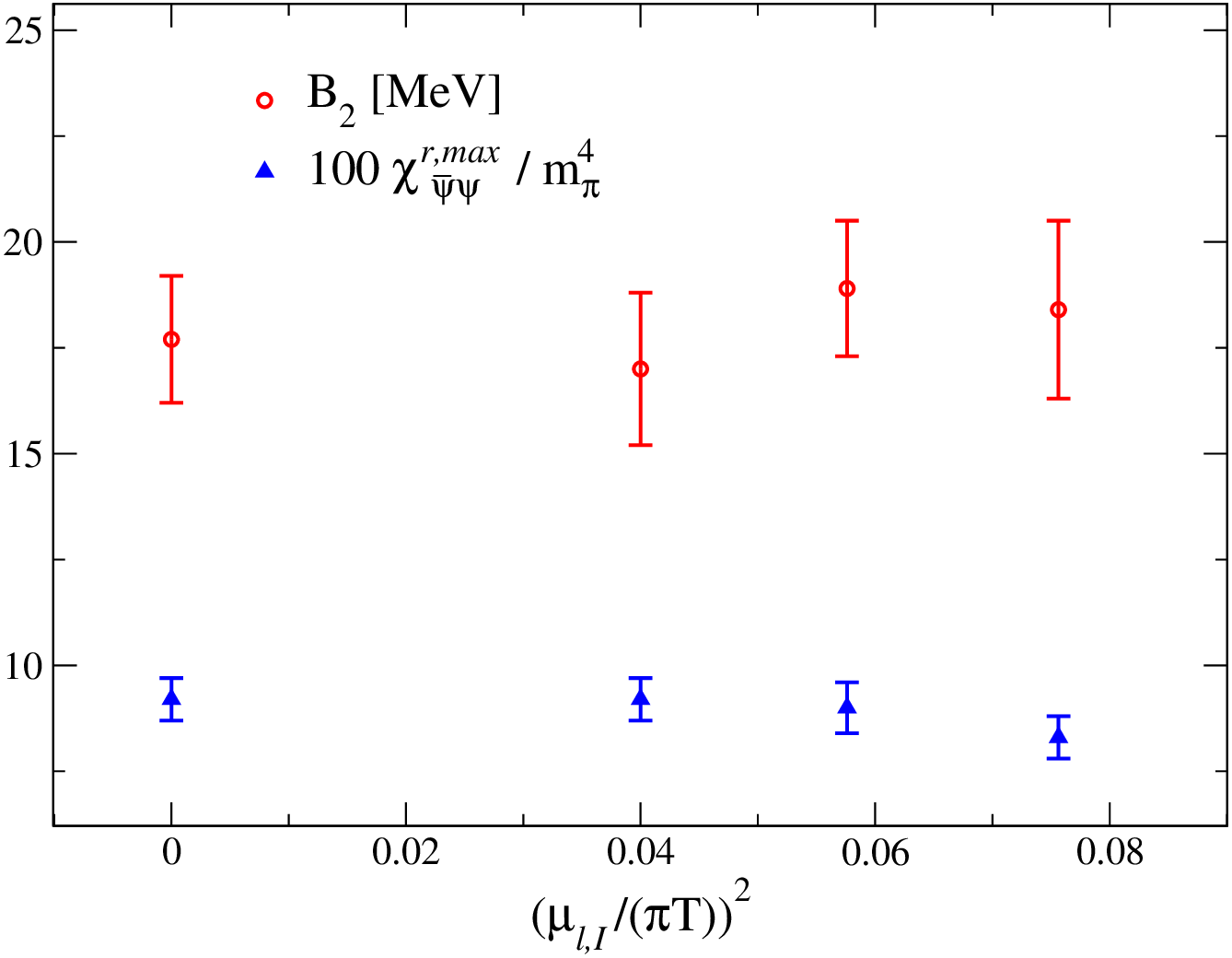}
\caption{Peak values ($\times 100$) and widths of the continuum extrapolated  
renormalized chiral susceptibility.}
\label{strength}
\end{figure}

To that purpose, in Fig.~\ref{strength} we plot the continuum extrapolated
width $B_2$ and height $A_2/B_2^2$ as a function of $\mu_{l,I}$. No apparent
change of either quantity can be appreciated, hence no dependence of the
strength as a function of $\mu_B$.

Of course, that does not exclude the presence of a critical endpoint at real
$\mu_B$: the critical region could be small enough, or the endpoint location
far enough from $\mu_B = 0$, so that no influence is visible for small,
imaginary $\mu_B$.  For instance, 
for $\mu_s = 0$, a Roberge-Weiss~\cite{RW} like endpoint is expected along the
pseudo-critical line at imaginary chemical potential, for $\mu_{l,I}/(\pi T)
\sim 0.45$~\cite{crow}. Fig.~\ref{strength} shows that also this endpoint has
no apparent influence on the strength of the transition in the explored range.

\subsection{Inclusion of $\mu_s \neq 0$ and systematics of
analytic continuation}
\label{mu_mu}

\begin{figure}[t!]
\includegraphics[width=0.92\columnwidth, clip]{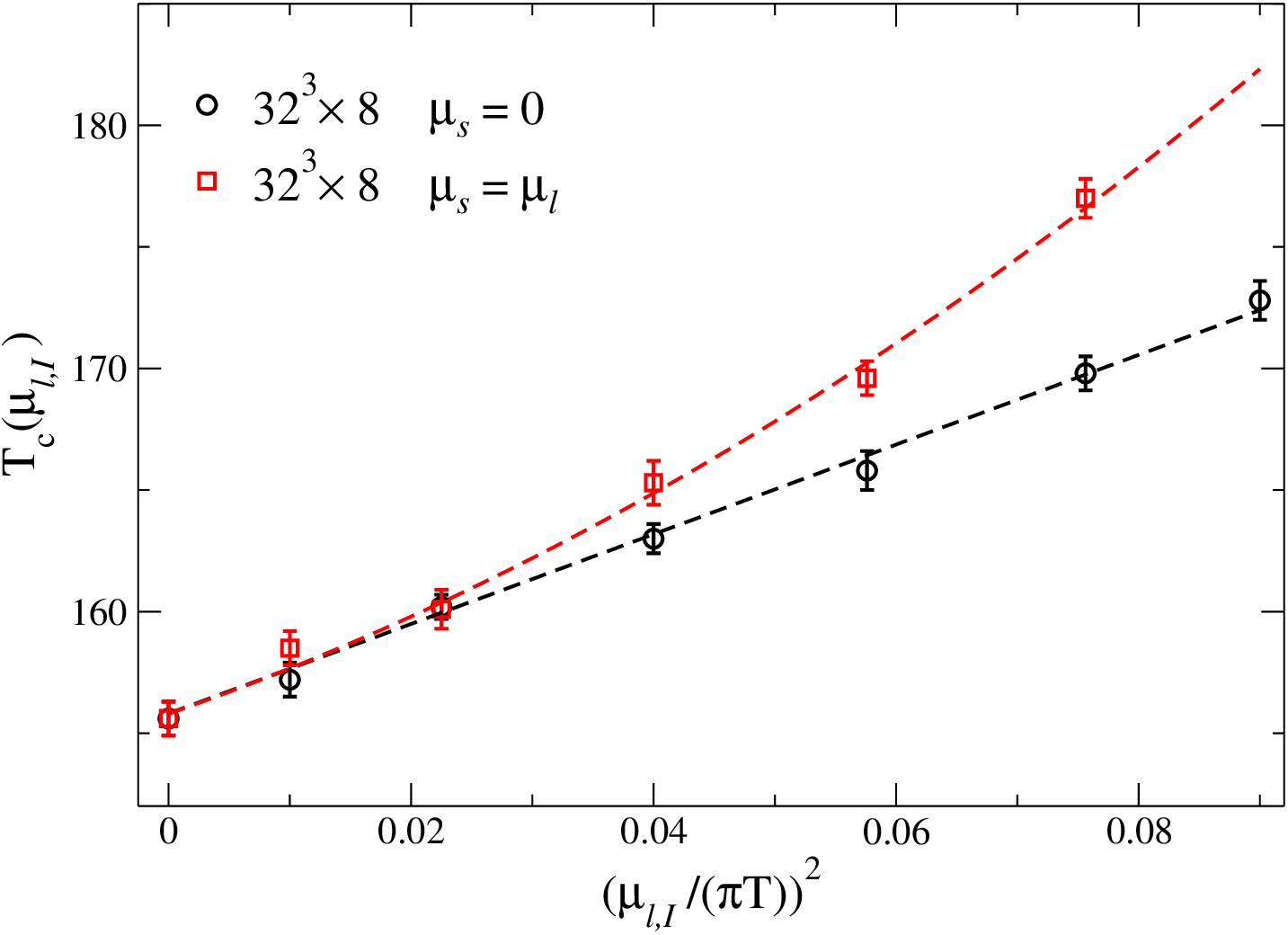}
\caption{Critical lines for the $32^3\times 8$ lattices in the two different 
setups: $\mu_s=0$ and $\mu_s=\mu_{l}$.}
\label{common_fit}
\end{figure}

We have extended results for $N_t = 8$ presented in Ref.~\cite{crow},
performing numerical simulations for a larger range of imaginary chemical
potentials, which include also the case $\mu_s = \mu_l$.  That enable us to
answer two important questions. What is the systematic error, in the
determination of $\kappa$ by analytic continuation, related to the truncation
of the Taylor series in Eq.~(\ref{fitfun}) and to the chosen range of chemical
potentials? What is the impact of our  effective ignorance about the actual
value of $\mu_s$ corresponding to the thermal equilibrium conditions?  We are
going to discuss in detail only the determination of the pseudo-critical
temperature from the renormalized chiral susceptibility, however we stress that
similar conclusions are reached when one considers the renormalized chiral
condensate.  The corresponding pseudo-critical temperatures, taken from
Table~\ref{tab:tc_all}, are reported in Fig.~\ref{common_fit} for $\mu_s = 0$
and for $\mu_s = \mu_l$.  

\begin{figure}[h!]
\includegraphics[width=0.92\columnwidth, clip]{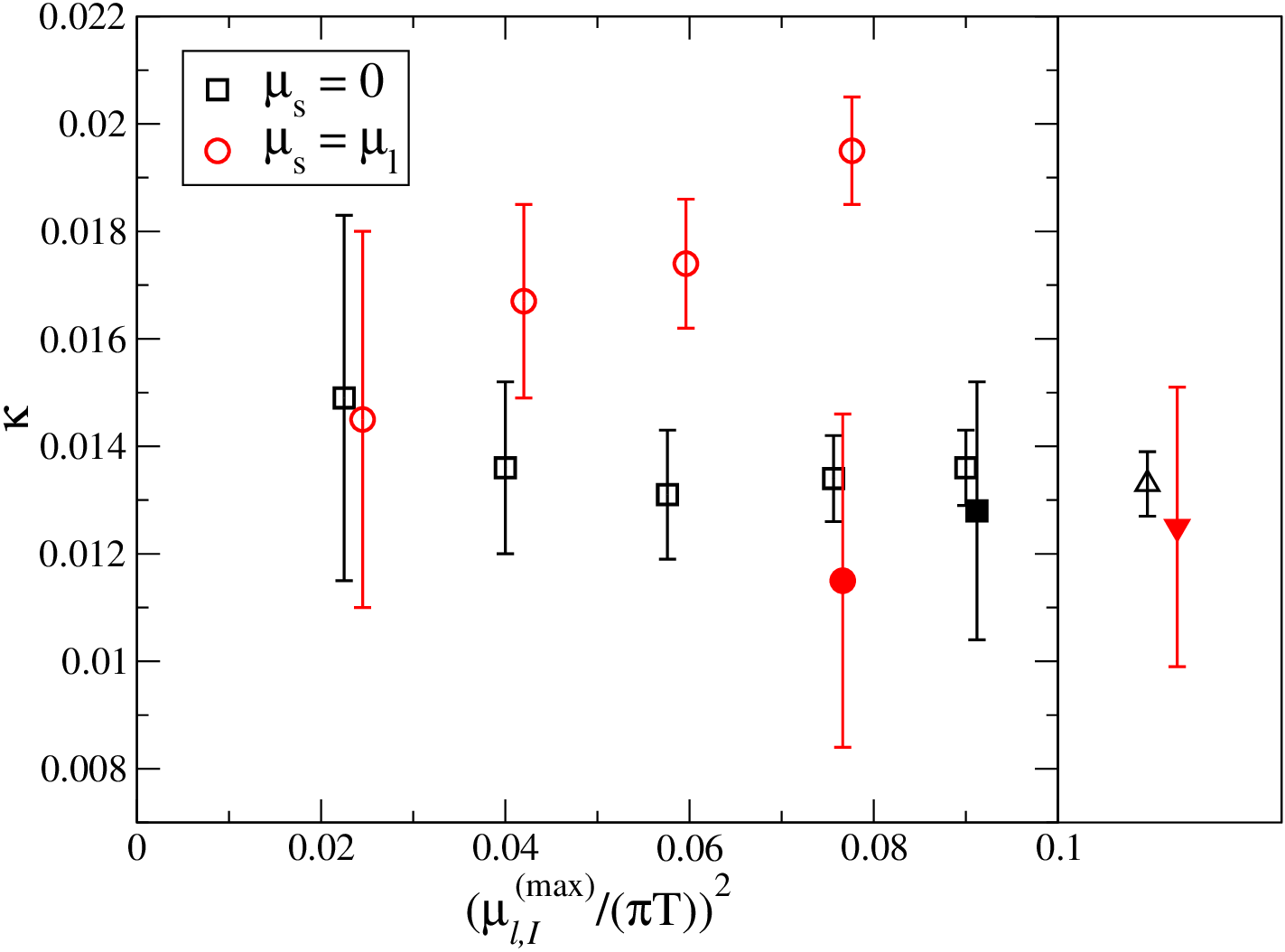}
\caption{Stability analysis of the fit to extract the $\kappa$ value for the 
        $32\times 8$ lattice. Empty 
        symbols correspond to purely quadratic fit while filled symbols 
        also take into account the quartic correction; red circles represents 
        the $\mu_s=\mu_l$ data, black squares the $\mu_s=0$ ones.  
        The right panel shows the result of a combined fit (i.e. fixing a common value for $T_c(0)$) to both data sets
        when a quartic correction is used for the $\mu_s=\mu_l$ data:
the open (filled) triangle corresponds to $\mu_s = 0$ ($\mu_s = \mu_l$).}
\label{range_fig}
\end{figure}

We first tried a quadratic fit in $\mu_{l,I}$: remembering the defintion
$\theta_l = \mu_{l,I}/T$, we used
\begin{equation}
T_c(\theta_{l}) = T_c(0) (1 + 9 \kappa \, \theta_l^2 )
\label{quad_fit}
\end{equation}
and several fits have been performed by changing each time the maximum value
$\mu_{l,I}^{(max)}$ included in the fit. Reasonable best fits are obtained in
all cases, apart from the fit to the whole $\mu_s = \mu_l$ range, which yields
a reduced $\tilde \chi^2 \sim 2.4$ and indicates the need for quartic
corrections in this case.  Results obtained for $\kappa$ are shown in
Fig.~\ref{range_fig}: for $\mu_s = 0$, the fitted value of $\kappa$ is
perfectly stable as the range of chemical potentials is changed. Instead, for
$\mu_s = \mu_l$, the value of $\kappa$ clearly depends on the fitted range of
chemical potentials: it is larger as the range is extended and becomes
compatible, within errors, with that obtained for $\mu_s = 0$ as the range is
decreased. This behavior is consistent with the presence of significant quartic
corrections in this case.  That may be related to the different structures of
the phase diagrams for imaginary chemical potential that one has in the two
cases: this issue has been discussed in detail in Ref.~\cite{crow}.

We then tried a best fit to a function including 
quartic corrections,
\begin{equation}
{T_c(\theta_{l})} = T_c(0) (1 + 9 \kappa \, \theta_l^2 
+ b \theta_l^4)\, ,
\end{equation}
to the whole range of chemical potentials explored in both cases. The
corresponding results obtained for $\kappa$ are reported in
Fig.~\ref{range_fig} as well. While for $\mu_s = 0$ the value is perfectly
compatible with the one obtained without including quartic corrections (indeed,
in this case one obtains $b = 0$ within errors), for $\mu_s = \mu_l$ we observe
a significant change, bringing $\kappa$ in good agreement with the $\mu_s = 0$
case.  A similar conclusion is reached when a common fit to both sets of data
(i.e. with a common value for $T_c(0)$) is performed, as shown in the right
panel of Fig.~\ref{range_fig} and in Fig.~\ref{common_fit}.

We conclude that, for $\mu_s = 0$, no evidence
of quartic corrections is found in the whole explored range.
As a consequence, the extracted $\kappa$ is stable against variations
of the fitted range and we can exclude the presence of significant
systematic corrections, related to the procedure of analytic continuation,
affecting the continuum extrapolated determination of $\kappa$ 
that we have provided.

In the case $\mu_s = \mu_l$, larger values of 
$\kappa$ are obtained when quartic corrections are neglected,
however $\kappa$ becomes compatible with that obtained for $\mu_s = 0$
when such corrections are included, or when the fitted range 
of chemical potentials is small enough. 
We conclude that $\kappa$ is not affected by the inclusion of 
$\mu_s$, at least within present errors, which however are larger
than for the $\mu_s = 0$ case. In particular, a fair estimate
in this case is 
$\kappa (\mu_s = \mu_l) = 0.013(3)$.

\section{Conclusions}
\label{sec_conclusions}

In the present study, we have extended results reported in Ref.~\cite{crow} by
performing numerical simulations on lattices with $N_t = 10,12$ and aspect
ratio $4$, and by enlarging the range of chemical potentials explored for $N_t
= 8$. That has permitted us to obtain continuum extrapolated results and to
better estimate possible systematics related to analytic continuation.

Regarding the case $\mu_s = 0$, we have obtained continuum extrapolated values
of $\kappa$ from different observables (chiral susceptibility and the chiral
condensate with two different renormalization prescriptions) and by two
different extrapolation procedures (extrapolating $\kappa_{cont}$ from $\kappa
(N_t)$ or extracting $\kappa_{cont}$ from continuum extrapolated temperatures).
{The comparison of the two different procedures permits us to give an 
estimate of the systematic uncertainties related to the continuum 
extrapolation. In the case of the renormalized chiral
susceptibility ($\kappa = 0.0132(10)$ vs $\kappa = 0.0131(12)$)
the systematic error is negligible in comparison to the statistical 
one. In the case of 
$\langle\bar{\psi}\psi\rangle^r_{(1)}$
($\kappa = 0.0134(13)$ vs $\kappa = 0.0145(11)$)
and of 
$\langle\bar{\psi}\psi\rangle^r_{(2)}$
($\kappa = 0.0127(14)$ vs $\kappa = 0.0138(10)$)
the systematic and statistical uncertainties are clearly comparable 
in size.
The extended analysis performed on  $N_t = 8$ has permitted us 
to state also that, within present errors,
 systematic effects
connected to the range of $\mu_l$ chosen to extract the curvature
are not significant.} Regarding finite size effects, the
analysis reported in Ref.~\cite{crow} already showed that they are negligible
within the present precision on lattices with aspect ratio 4.  Taking into
account the obtained results and the contributions from the
systematic effects mentioned above, we quote $\kappa = 0.0135(15)$ as our final
continuum estimate for the case $\mu_s = 0$. 

Such a result confirms, even after
continuum extrapolation, a discrepancy with previous determinations obtained by
Taylor expansion~\cite{Kaczmarek2011,Endrodi2011,Borsanyi2012}, reporting
$\kappa \sim 0.006$. {As already discussed quantitatively in 
Ref.~\cite{crow}, only part of this discrepancy can be accounted for by
the different prescriptions used to determine the 
dependence of $T_c$ on $\mu_l$. Contrary to the Taylor 
expansion case, when 
working  at imaginary $\mu_l$ one can use consistently the same 
prescription to locate $T_c$ used for $\mu_l = 0$, i.e.
looking for the maximum of the chiral susceptibility or 
the inflection point of the chiral condensate 
 (see Ref.~\cite{crow} for more details).
The remaining part of the discrepancy could be possibly attributed 
to the systematic uncertainties related
to the continuum extrapolation of previous studies.
However, we stress that updated investigations by
the same groups lead to 
results which are consistent with our estimate
(see, e.g., Ref.~\cite{hedge}).
}

Regarding the case $\mu_s = \mu_l$, we have confirmed the preliminary results
reported in Ref.~\cite{crow}.  There is evidence for the presence of quartic
contributions in the dependence of $T_c$ on the imaginary $\mu_B$ in this case
and when such contributions are taken into account, or when the range of fitted
chemical potentials around $\mu_B = 0$ is small enough, the curvature becomes
compatible, even if within larger errors, with that obtained for $\mu_s = 0$.
That means that also for the equilibrium conditions created in heavy ion
collisions, corresponding to $\mu_s \sim 0.25\, \mu_l$ around $T_c$, one does
not expect significant deviations from the results obtained for $\mu_s = 0$: a
prudential estimate for the curvature in this case is\footnote{{After
completion of this work, Ref.~\cite{ntc} has appeared, 
reporting the consistent result $\kappa = 0.0149(21)$.}}
$\kappa = 0.0135(20)$.
That is obtained based on the estimate for $\mu_s = 0$, with an increased error
determined on the basis of the uncertainty that we have for the curvature
extracted at $\mu_s = \mu_l$.

Finally, the analysis of the continuum extrapolated peak of the chiral
susceptibility as a function of imaginary $\mu_B$ shows no significant
varations of the strength of the transition, which could be associated to a
possible nearby critical endpoint present along the pseudo-critical line.

\acknowledgments

\begin{table}[t!]
\centering
\begin{tabular}{ |c|c|c|c|c| }
\hline
$\beta$ & Lattice & $\chi_{\bar{\psi}\psi}$ & \rule{0mm}{3.5mm} $\langle\bar{\psi}\psi\rangle - 2(m_l/m_s)\langle\bar{s}s\rangle$ & $\langle\bar{\psi}\psi\rangle/2$ \\
\hline
3.50 & $32^4$                 & 1.97(4)  & 0.07999(11) & 0.04403(5) \\
3.55 & $32^4$                 & 1.97(5)  & 0.05680(13) & 0.03164(7) \\
3.60 & $32^4$                 & 2.05(6)  & 0.03912(14) & 0.02211(7) \\
3.65 & $32^4$                 & 1.82(3)  & 0.02633(2)  & 0.01518(9) \\
3.70 & $32^4$                 & 1.80(3)  & 0.01804(3)  & 0.01064(2) \\
\hline
3.65 & $48^4$                 & 1.74(7)  & 0.02638(4)  & 0.01521(2) \\
3.75 & $48^4$                 & 1.61(5)  & 0.01232(5)  & 0.00749(2) \\
3.85 & $48^4$                 & 1.47(4)  & 0.00614(2)  & 0.00401(1) \\
3.95 & $48^4$                 & 1.37(3)  & 0.00331(2)  & 0.00237(1) \\
\hline
\end{tabular}
\caption{Determination of the observables at $T=0$ (on the $32^4$ and $48^4$ lattices)
needed to perform the renormalizations discussed in Section~\ref{sec2}. Data are in lattice units.}
\label{tab:simT0}
\end{table}

FS received funding from the European Research Council under the European
Community Seventh Framework Programme (FP7/2007-2013) ERC grant agreement No
279757.  FN acknowledges financial support from the INFN SUMA project.
Simulations have been performed on the BlueGene/Q Fermi at CINECA (Projects
Iscra-B/EPDISIM, Iscra-B/CROWQCD and INF14\_npqcd), and on the CSN4 Zefiro
cluster of the Scientific Computing Center at INFN-PISA.

\appendix

\section{{Data at $T=0$}}\label{app:t0}

The determination of the renormalized condensate and susceptibility requires
the computation of the corresponding quantities at $T = 0$ and at the same UV
cutoff of the finite temperature data.  To that aim, we spanned a range of
$\beta$ on the line of constant physics, $3.5 \leq \beta \leq 3.95$. The
lattice sizes have been chosen in such a way to have temperatures well below $T_c$, keeping at
the same time finite size effects under control.  This required us to perform
simulations on larger lattices (going from $32^4$ up to $48^4$) as we decreased
the value of the lattice spacing.  We report results in Table~\ref{tab:simT0}.

\begin{figure}[h!]
\includegraphics[width=0.92\columnwidth, clip]{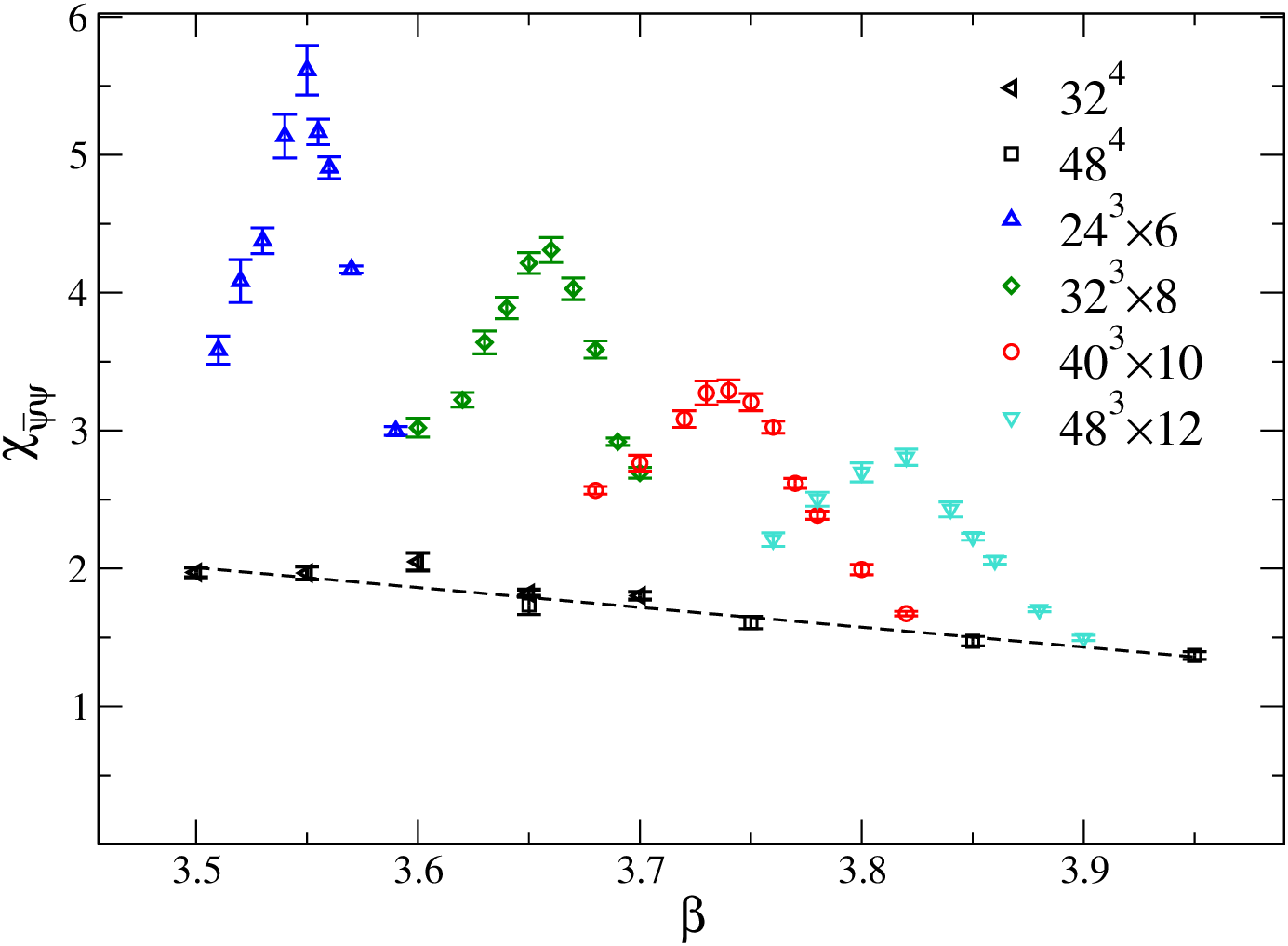}
\caption{Comparison between the $T = 0$ and $T \neq 0$ chiral susceptibility
$\mu_B = 0$. The $T=0$ susceptibility is needed to compute the renormalized chiral
susceptibility, Eq.~(\ref{rensusc}).  Data are in lattice units and a linear
fit to the $T = 0$ data is shown.}
\label{susc_T0}
\end{figure}

The temperatures, which are in the range $T\sim \ 25 - 50$ MeV, are low enough
to be considered as a good approximation of the $T = 0$ limit; indeed, as
expected because of the absence of transitions in this $T$ range, observables
depend smoothly on $\beta$; moreover no dependence at all is expected on the
imaginary chemical potentials, since they can be viewed as a modification in
the temporal boundary conditions which, at $T = 0$ (i.e.  for infinite temporal
extension), are completely irrelevant.  Hence, the relatively coarse sampling
of the interval is enough to permit a reliable interpolation.  We adopted a
cubic spline interpolation for the condensate and a linar fit for the
susceptibility.

The renormalization prescription for the susceptibility, Eq.~(\ref{rensusc}),
requires the subtraction of the $T=0$ result from the finite $T$ contribution.
To give an idea of the relative magnitude of this subtraction, in
Fig.~\ref{susc_T0} we plot $\chi_{\bar{\psi}\psi}$ for zero chemical potential
and both at zero and finite $T$.

\end{document}